\documentclass[twocolumn,twocolappendix]{aastex63}
\usepackage{amsmath}

\newcommand\teff{$T_\mathrm{eff}$}

\newcommand\logt{\log T_{\mathrm{eff}}}

\newcommand\msun{M$_\odot$}
\newcommand\rsun{R$_\odot$}

\defcitealias{kounkel2019a}{Paper I}
\defcitealias{kounkel2020}{Paper II}
\defcitealias{mcbride2021}{Paper III}
\usepackage{listings}

\begin{document}
\title{Untangling the Galaxy. IV. Empirical Constraints on Angular Momentum Evolution and Gyrochronology for Young Stars in the Field}

\author[0000-0002-5365-1267]{Marina Kounkel}
\affiliation{Department of Physics and Astronomy, Vanderbilt University, Nashville, TN 37235, USA} 
\email{marina.kounkel@vanderbilt.edu}
\author[0000-0002-3481-9052]{Keivan G.\ Stassun}
\affiliation{Department of Physics and Astronomy, Vanderbilt University, Nashville, TN 37235, USA}
\author[0000-0002-0514-5538]{Luke G. Bouma}
\altaffiliation{51 Pegasi b Fellow}
\affiliation{Cahill Center for Astronomy and Astrophysics, California Institute of Technology, Pasadena, CA 91125, USA}
\author[0000-0001-6914-7797]{Kevin Covey}
\affiliation{Department of Physics and Astronomy, Western Washington University, 516 High St, Bellingham, WA 98225, USA}
\author{Lynne A. Hillenbrand}
\affiliation{Department of Astronomy, California Institute of Technology, Pasadena, CA 91125, USA}

\newcommand{\columbia}{Department of Astronomy, Columbia University, 550 West 120th Street, New York, NY 10027, USA}
\author[0000-0002-2792-134X]{Jason Lee Curtis}
\affiliation{\columbia}

\begin{abstract}
We present a catalog of $\sim$100,000 periodic variable stars in TESS FFI data among members of widely distributed moving groups identified with Gaia in the previous papers in the series. 
By combining the periods from our catalog attributable to rotation with previously derived rotation periods for benchmark open clusters, we develop an empirical gyrochronology relation of angular momentum evolution that is valid for stars with ages 10--1000~Myr. Excluding stars rotating faster than 2~days, which we find are predominantly binaries, we achieve a typical age precision of $\approx$0.2--0.3~dex and improving at older ages. 
Importantly, these empirical relations apply to not only FGK-type stars but also M-type stars, due to the angular momentum distribution being much smoother, simpler, continuous and monotonic as compared to the rotation period distribution. As a result, we are also able to begin tracing in fine detail the nature of angular momentum loss in low-mass stars as functions of mass and age. 
We characterize the stellar variability amplitudes of the cool stars as functions of mass and age, which may correlate with the starspot covering fractions. 
We also identify pulsating variables among the hotter stars in the catalog, including $\delta$~Scuti, $\gamma$~Dor and SPB-type variables. 
These data represent an important step forward in being able to estimate precise ages of FGK- and M-type stars in the field, starting as early as the pre-main-sequence phase of evolution. \\
\end{abstract}


\section{Introduction}

Stellar ages are fundamental calibrators for our understanding of star formation, planet evolution, and the evolution of the Milky Way. Generally speaking though, it easiest to measure accurate and precise ages using ensembles of identically-aged stars, rather than lone field stars.
In recent years, parallaxes and proper motions measured by the Gaia satellite \citep{gaia-collaboration2018} have enabled the identification of thousands of such stellar groups. 
In particular, \citet[hereafter, Paper I]{kounkel2019} and \citet[hereafter, Paper II]{kounkel2020} identified almost 1 million stars that are candidate members of more than 8,000 co-moving groups within $\sim$3 kpc.
These structures include many previously known open clusters and moving groups, and they are contiguous in 3D position and 2D velocity space.
The ages of these populations can be estimated through isochrone fitting.


However, over time, stellar groups lose their memory of the initial kinematics with which they formed. Increasingly fewer and fewer stars can be identified as members of the moving groups at older ages, and populations (especially the ones that originally had relatively few members) will eventually entirely dissolve into the Galaxy. After this happens, other methods are needed to determine their age.

Ages of low mass stars up to several tens of million years (Myr) can be measured via comparisons to isochrones in photometric or spectroscopic HR diagrams if they are pre-main sequence \citep{marigo2017,mcbride2021}. However, solar type stars quickly reach the main sequence and are not susceptible to isochrone dating after that point, until they near the end of their main-sequence lifetime. 

Gyrochronology offers an alternative opportunity. 
As a star gets older, magnetic braking slows its rotation citep{weber1967,skumanich1972}. 
If the star has inhomogeneities such as starspots or faculae on its surface, it can be possible to measure the star's rotation period and relate it to stellar age \citep{barnes2003,barnes2007,barnes2010}. With the advent of all-sky surveys that provide detailed light curves for a large number of stars, such as 
the Transiting Exoplanet Survey Satellite \citep[TESS;][]{ricker2015}, 
applying gyrochronology relations to a large number of stars to determine ages is becoming increasingly more feasible. This technique is particularly effective for solar-type stars, due to the most pronounced evolution of periods in this mass range.

However, a major limitation is a lack of sufficiently accurate gyrochrone models. While theoretical gyrochrone models have made substantial progress \citep[e.g.,][]{Matt2015,spada2020,gossage2021}, they still do not appear to fully reproduce the observed features of the empirical period--temperature relations across all ages and masses \citep{curtis2020}, and often require parameters that are difficult to measure empirically. The current state-of-the-art method of estimating stellar ages through gyrochronology is through comparing rotation periods of a young population to the rotation periods of a handful of pristine open clusters to confirm similarity \citep[e.g.,][]{curtis2019,andrews2022, bouma2021}. However, this method does not allow the precise interpolation between the calibrating clusters' ages due to limited number of clusters for which rotation periods have been measured. \citet{angus2019} have created empirical isochrones that have been calibrated to Praesepe, but, although they perform well on FG-type stars in older populations \citep{angus2020}, they significantly underestimate ages for populations younger than $\sim$500 Myr \citep{curtis2020}. TESS is well positioned to provide rotation periods that could be used to anchor a new, better empirical gyrochronology model for stars in this age range, as the 28 day duration of its light curves provide the maximum sensitivity to periods shorter than $\sim$10 days (corresponding to the rotation periods of F, G, and M stars younger than Praesepe).

In this work we develop an empirical gyrochrone fit using the catalog of sources presented in \citetalias{kounkel2020}. This fit is based on populations that are near-continuous in their age distribution, and extends in age up to 1 Gyr. In Section \ref{sec:data} we assemble the catalog of rotation periods for the stars with known ages using TESS FFIs. In Section \ref{sec:results} we describe the properties of the stars with convective envelopes that are observed in these data, as well as develop the empirical relation of stellar angular momentum evolution. In Section \ref{sec:discussion} we discuss the potential of the gyrochrone relations to determine ages of the field stars. In Section \ref{sec:conclusion} we summarize and conclude this work.

\section{Data} \label{sec:data}

\subsection{Selection of stars with ages from Theia catalog}\label{sec:theia}

In \citetalias{kounkel2020} we identified stellar associations and moving groups in Gaia DR2 \citep{gaia-collaboration2018} using hierarchical clustering with HDBSCAN \citep{hdbscan1}. The clustering was performed in 5D phase space, including $l$, $b$, parallax, as well as tangential velocities. An average age of all the moving groups was determined using Gaia and 2MASS photometery using a data-driven tool Auriga that performs pseudo-isochrone fitting through interpolating photometry of a variety of clusters with known ages.

The final catalog (collectively referred to as Theia), consisted of 8293 moving groups within 3 kpc. These groups, ranging in age from $<$10 Myr to $>$1 Gyr, span a large variety of scales, consisting from 40 to 10,000 stars, some concentrated in compact clusters, some spanning long strings extending upwards of 200~pc. The typical precision in age in this catalog is $\sim$0.1 dex, reaching $\sim$0.2 dex in some of the sparser and more evolved groups where the precise location of the main sequence turn-off is uncertain.

To extract a sample of nearby stars with good prospects for TESS rotation period measurements, we selected sources within 500~pc and that have $T<16$ mag, which is the typical magnitude limit for $\sim$1\% precision in fluxes with TESS. We adopted $T$ band magnitudes from the TESS Input Catalog \citep{stassun2019}. Additionally we included all the sources for which 2-minute cadence target pixel data was available (typically $T<13$ mag), regardless of their distance. 

Although the catalog does include sources as young as a few Myr, such sources exhibit complex, aperiodic variability \citep{cody2014} that will compete with the stellar rotation signals of primary interest in this work. To avoid this complication, we limit the sample to primarily focus on $>$10 Myr based on the ages provided in the Theia catalog, and defer a more comprehensive discussion of the younger sample to a subsequent work.

\subsection{Matching to TESS and processing of light curves}\label{sec:tess}

We matched the sources from the Theia catalog to the footprint of TESS in the first 35 sectors. This upper limit was set by the available data at the time of our analysis. The first 26 sectors covered 70\% of the sky, excluding a $\sim15^\circ$ gap around the ecliptic, and also excluding a $140^\circ \times 20^\circ$ region north of the galactic center that would have been heavily affected by scattered light from the Earth and Moon had it been observed.
Sectors 27--35 roughly repeated the coverage of the first nine sectors. We used the \texttt{eleanor} package \citep{feinstein2019} to download light curves for the 102,112 Theia sources that were observed in the TESS full frame images of sectors 1-35. As all of the sectors were processed separately, this resulted in 226,003 light curves\footnote{We have made them available at \url{https://doi.org/10.5281/zenodo.6757610}}.

We have attempted to use other light curve generation techniques, such as those from \citet{huang2020} or those generated through TESS SIP \citep{hedges2020}. However, the former was not sensitive to rotation periods longer than 2 days due to detrending routines used to generate those light curves, and the latter introduced noise, particularly in form of an artificial $\sim$90 day period (which was particularly pronounced in stars in the continuous viewing zone due to abundance of longer baseline data) that has made real periods of a few days difficult to detect. Of these, \texttt{eleanor} provided the most robust light curves to all periods $<$12 days. Unfortunately, longer periods cannot be effectively recovered with it at this time. As each 28 day sector consists of two TESS orbits, $\sim$14 days each, there are strong systematics that introduce periodicity comparable to the duration of an orbit. Correcting these systematics preferentially suppresses long periods of true variable stars as well.

For the specific problem of detecting signals with periods longer than ten days, TESS is a systematics-limited instrument. Data quality issues most often related to scattered light from the Earth and Moon near the time of spacecraft perigee passes can yield artificial spikes and drops in the light curves. To automatically exclude light curve cadences that are problematic, we first normalize data in each sector by the median, and limit the flux to be within 50\% of the median (i.e, normalized flux in the range of 0.5--1.5). We then further limit the flux to be within three standard deviations of the flux found within the remaining epochs within the sector; this step is repeated four times. This excludes the vast majority of artificial data issues. If a source is an eclipsing binary, typically the eclipses are also deleted from the light curve, leaving only the continuum (from which the rotational signature is more cleanly measured). However, contact binaries may remain in the sample. Additionally, close binaries can have stellar rotation periods that are tidally synchronized to their orbital periods. Such sources would be retained in the sample.

\subsection{Measurement of periods from TESS light curves}\label{sec:periods}

We measure periodic signals using Lomb--Scargle periodograms for each source in each individual sector. Stitching multiple sectors together, unfortunately, is not effective in recovering periods $>$12 days. Additionally, short periods that can be detected within a single sector can evolve over time, producing a slightly different period, or the periodic signal can completely disappear into the noise due to a decrease in the amplitude. This can happen due to the evolution of spots on the stellar surface. 

\begin{figure*}
\epsscale{1.1}
\plotone{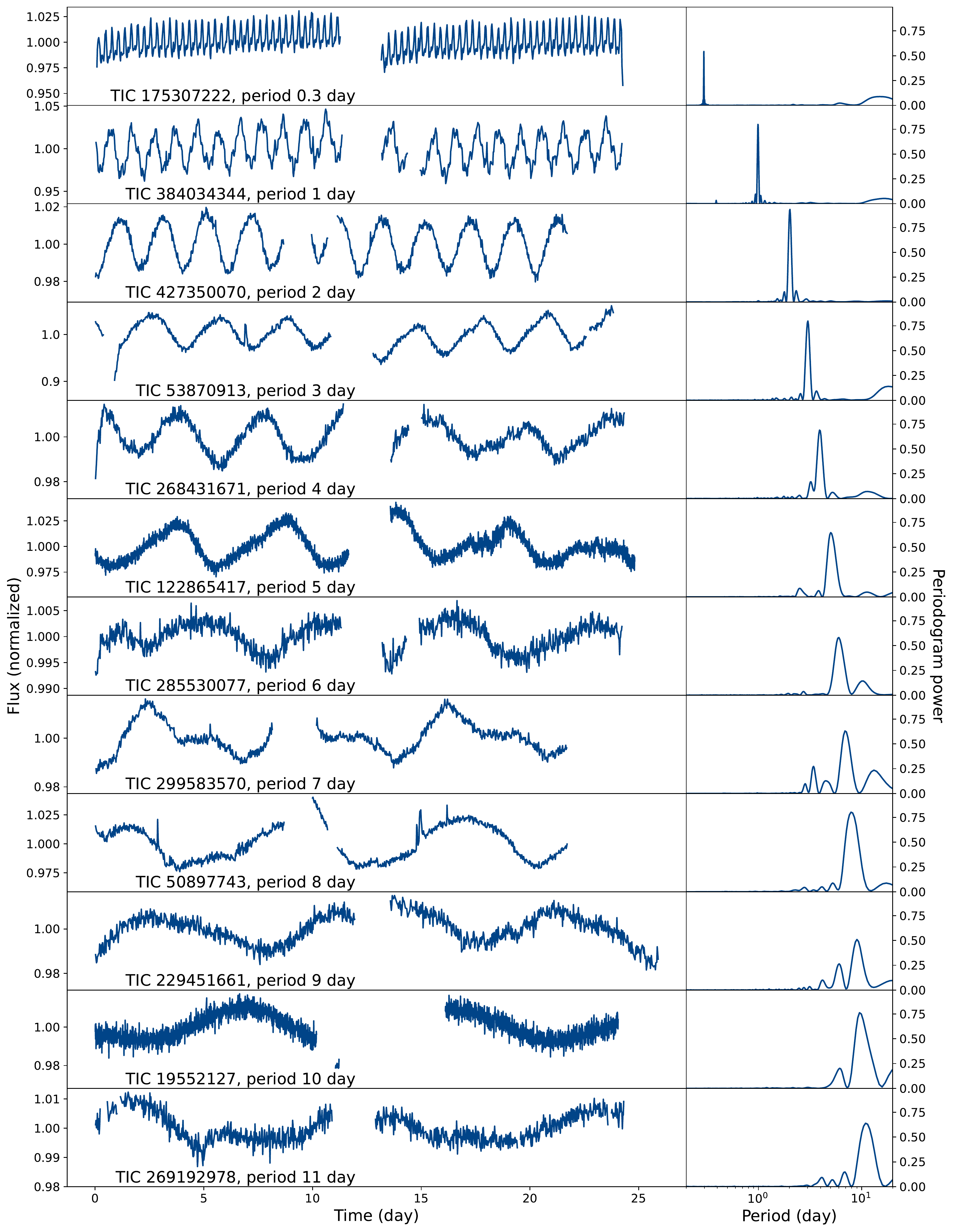}
\caption{Example of the periodic light curves and the corresponding Lomb--Scargle periodograms of the sources in this work.
\label{fig:lcs_ex}}
\end{figure*}

The periodic signal can often be somewhat weak; furthermore, there may be multiple peaks in the periodogram. Due to the number of sources, it is difficult to visually examine all sources to select those with only a single dominant period. We use \texttt{GaussPy} to perform an Autonomous Gaussian Deconvolution \citep{lindner2015,lindner2019} to identify all possible signals. The periodogram is created using the \texttt{astropy} timeseries package, using an automatically generated list of periods from 4 hours to 33 days, with 10 samples per typical peak, such that the sampling is higher at higher frequencies. Separately we also generate a periodogram from 40 minutes to 4 hours to look for rapid variability such as that attributable to $\delta$ Scuti variables, but without affecting the ability of recovering long period signal.

We then fit the periodogram in pixel space, resulting in all peaks having an approximately similar width. We set the \texttt{GaussPy} $\alpha$ parameter (which controls the typical width of identified Gaussians in the data) to 0.5, and subtracting an amplitude of 0.01 from the periodogram to ignore insignificant peaks. Following the identification of all the significant periodogram peaks, we interpolate their pixel mean location back to the period space $P$, which we record alongside with the amplitude $A$ of the Gaussians.

\texttt{eleanor} allows light curves to be generated with different levels of processing and detrending. One of the outputs, CORR\_FLUX, is a light curve that is corrected for various known systematic effects on an orbit-by-orbit basis. These CORR\_FLUX outputs appear to be the most stable for the vast majority of the light curves of all intensities for periods $<$8 days (or, approximately, on the timescales of half of an orbit of TESS). However, the CORR\_FLUX processing completely suppresses longer period signals. As such, to compensate, we independently process periodograms for PCA\_FLUX (light curve generated via subtraction of co-trending basis vectors derived via the principle component analysis) and RAW\_FLUX (light curve without any processing applied). Comparing the derived periods to the periods measured by other surveys such as ASAS-SN \citep{jayasinghe2018}, we devise the set of criteria to most faithfully recover the underlying periodic signal.

\begin{figure}[!ht]
\includegraphics[width=\linewidth]{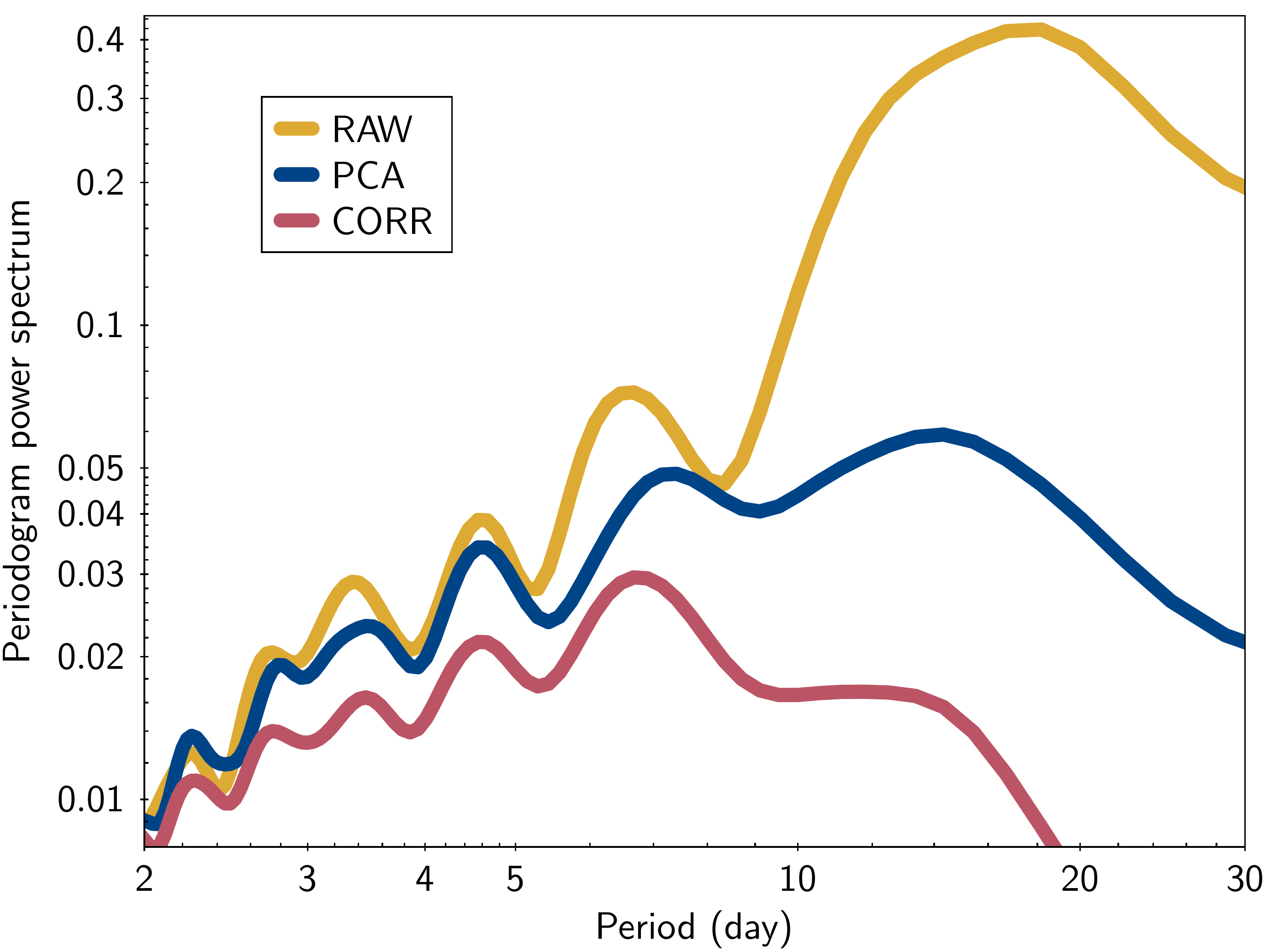}
\caption{An average periodogram of 7431 CORR, PCA, and RAW lightcurves for stars with \teff\ between 8000 and 9000 K that appear to be aperiodic. This highlights the artificially injected periods due to the systematics from TESS, most notably at $\sim$7 days and 14 days, corresponding to 0.5 and 1 orbit of TESS around the Earth. RAW light curves are most susceptible to these systematics but can recover the real periods $>$7 days most faithfully compared to the literature. CORR light curves are least dominated by systematics, but they largely suppress longer periods in their corrections.
\label{fig:power}}
\end{figure}

\begin{enumerate}
 \item We record a primary period from the CORR\_FLUX periodogram for a star if the primary amplitude $A_{1,CORR}$ is higher than $>0.3$. Additionally, we also require the large difference in ratios between primary and secondary amplitudes $A_{2,CORR}/A_{1,CORR}<0.7$ for sources with period $>$0.2 days. Sources with shorter ($<$ 0.2d) periods tend to be multi-periodic---not only intrinsically, but also due to the minimum cadence of TESS FFIs introducing a beat frequency. We record their primary period as is if they meet the amplitude check, even if they fail the ratio check.
 \item For the remaining sources we consider other alternatives. PCA\_FLUX can be somewhat more sensitive to the longer periods, but it can be more susceptible to noise (Figure \ref{fig:power}). As such, we require $A_{1,PCA}>0.45$, $P_{1,PCA}<11$ days, as well as, similar to the above $A_{2,PCA}/A_{1,PCA}<0.7$ or $P_{1,PCA}<0.2$ day.
 \item Raw light curves can faithfully preserve long periods. They are also prone to the failure case of primarily detecting signal that is comparable to the orbital period of TESS (Figure \ref{fig:power}). However, generally, when the signal is real, there is also sufficient power in PCA periodogram, even though the period in PCA light curve may be less inconsistent with literature than the period from the raw light curve. As such, for the remaining sources we record $P_{1,RAW}$ if $A_{1,PCA}>0.3$ (not $A_{1,RAW}$). However, we similarly require $A_{2,RAW}/A_{1,RAW}<0.7$ or $P_{1,RAW}<0.2$ day. The maximum period we record is $P_{1,RAW}<12$ days.
 \item Furthermore, we retain the primary period regardless of its amplitude in the periodogram, if it is within 5\% in all three lightcurves for the same source in the same sector.
\end{enumerate}

\subsection{Evaluation of TESS period recovery}

TESS has a low spatial resolution, with each pixel having a size of 21$''$. Because of this, contamination from other stars along a similar line of sight, particularly along the Galactic plane and/or within dense clusters, can often be a concern for this instrument. \texttt{eleanor} has the ability to automatically select the most appropriate aperture for the star to integrate over, depending on its brightness and density around it, to attempt to minimize this contamination. This can make it difficult to determine what aperture was used, and how many other sources may contribute to the flux within that light curve.

We compute the flux contamination ratio, defined as the total flux of all stars in Gaia in a given radius divided by the flux of a target star, to test the effect of potential contamination. We check 6 different radii, in steps of 10$''$, from 10$''$ (i.e., radius of a pixel) to 60$''$ (exceeding the size of a typical aperture). We also record the renormalised unit weight error (RUWE) from Gaia to identify likely close binaries which can also lead to contamination. We note that the results presented in Section~\ref{sec:results} are robust against any cuts in either RUWE or the flux ratio, but applying these cuts preferentially excludes lower mass stars, as such, in the interest of retaining maximum sensitivity to all masses we do not apply any filtering.

\begin{figure*}[!ht]
\includegraphics[width=0.5\linewidth]{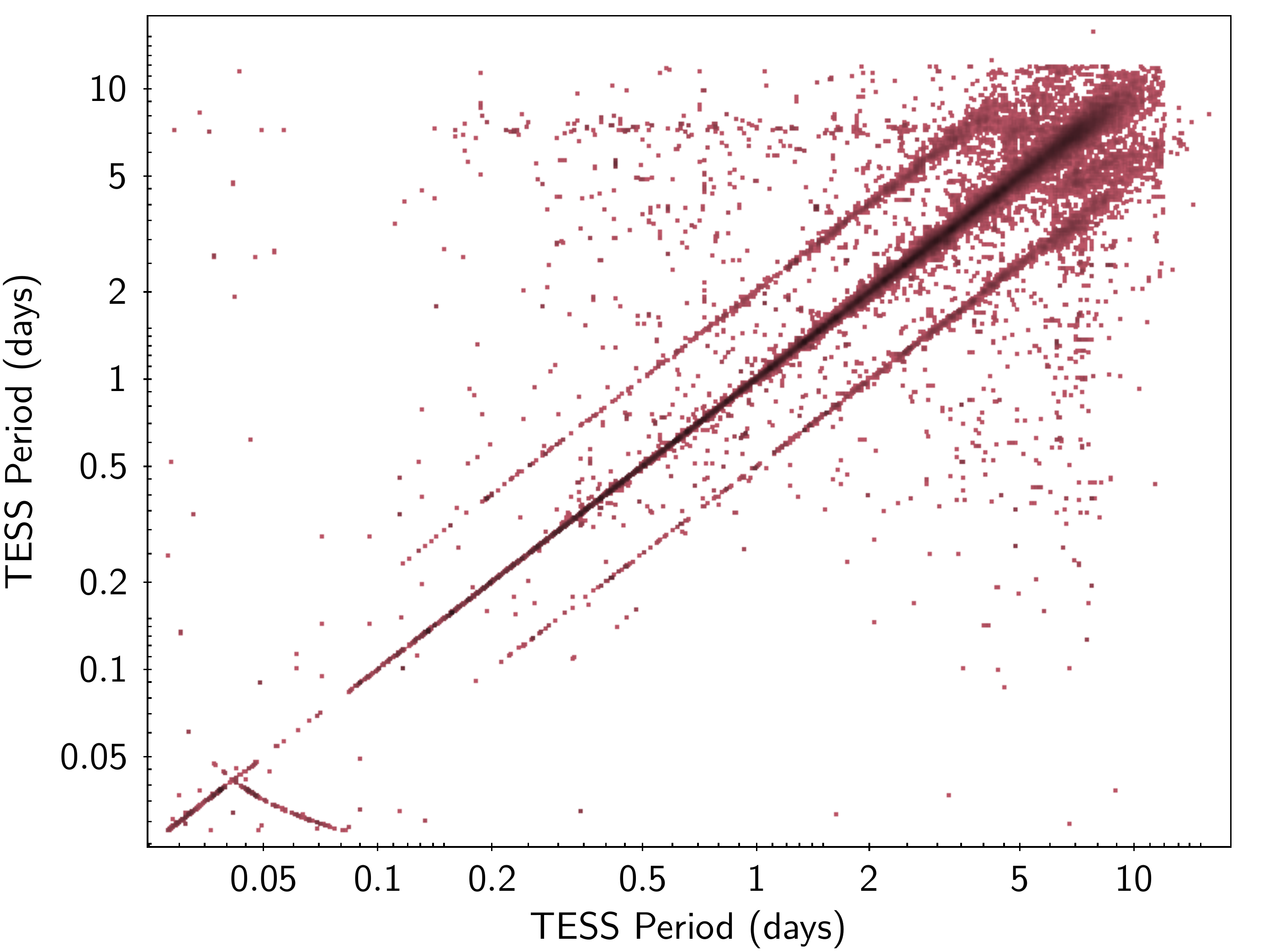}
\includegraphics[width=0.5\linewidth]{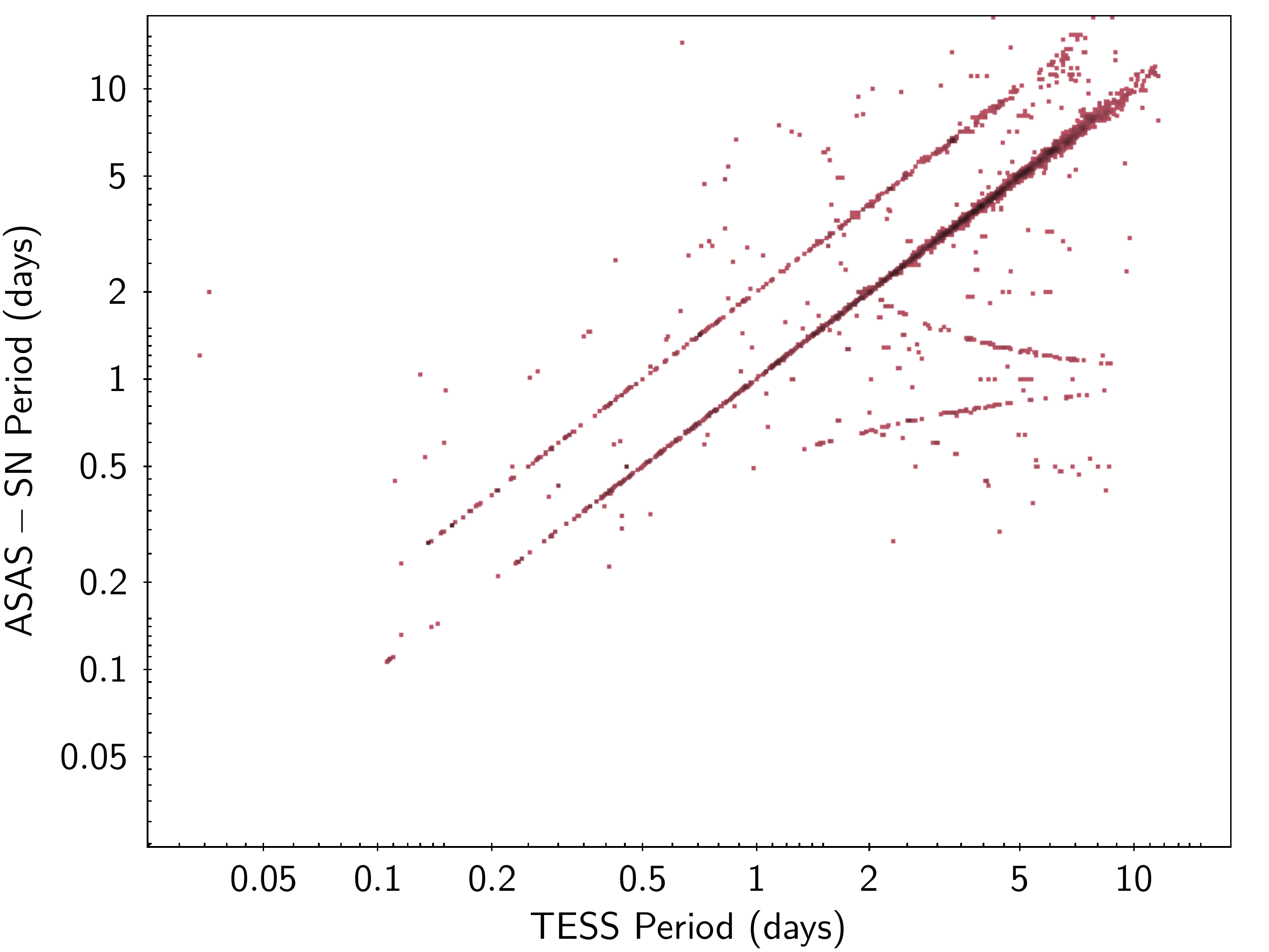}
\includegraphics[width=0.5\linewidth]{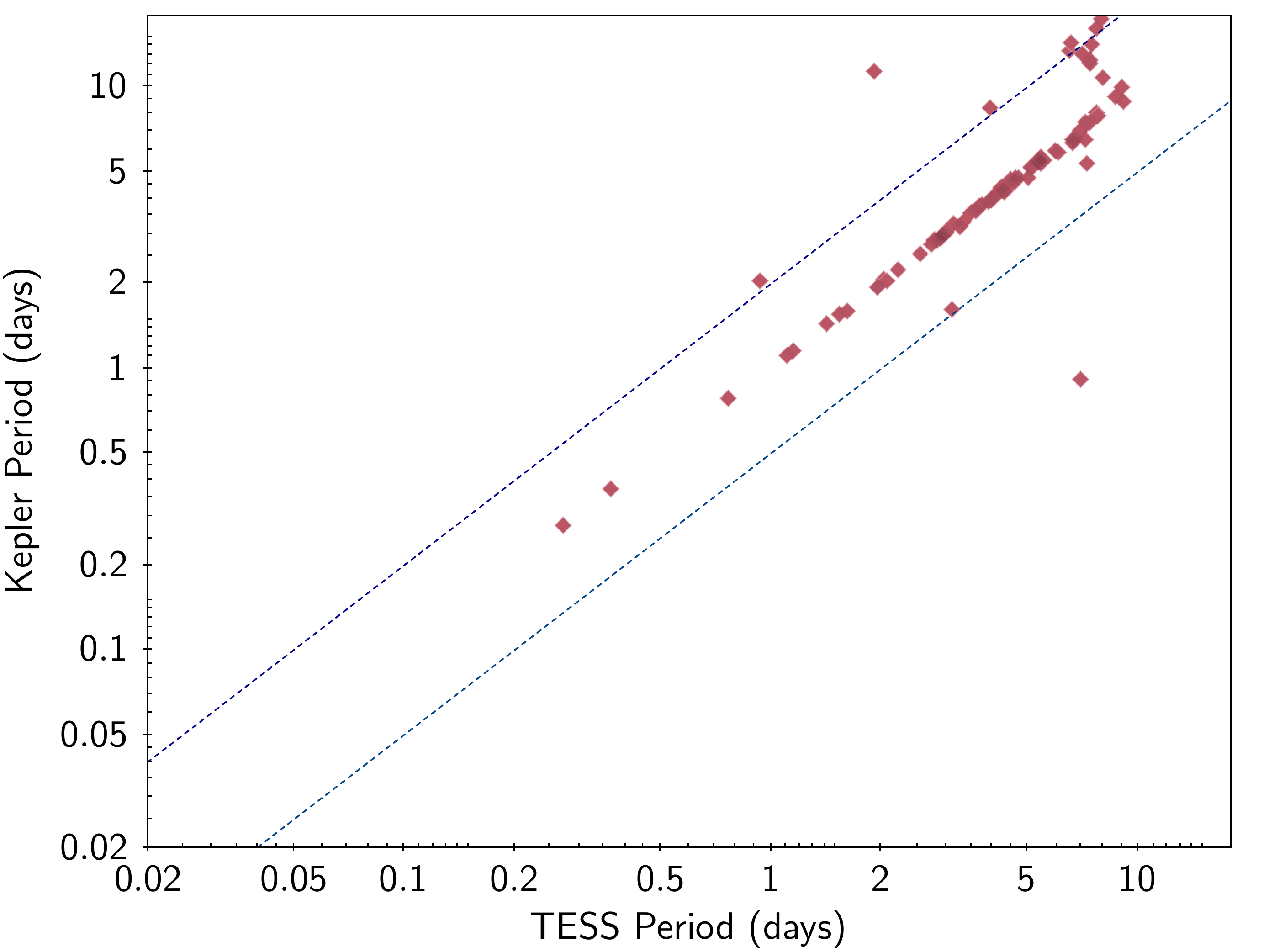}
\includegraphics[width=0.5\linewidth]{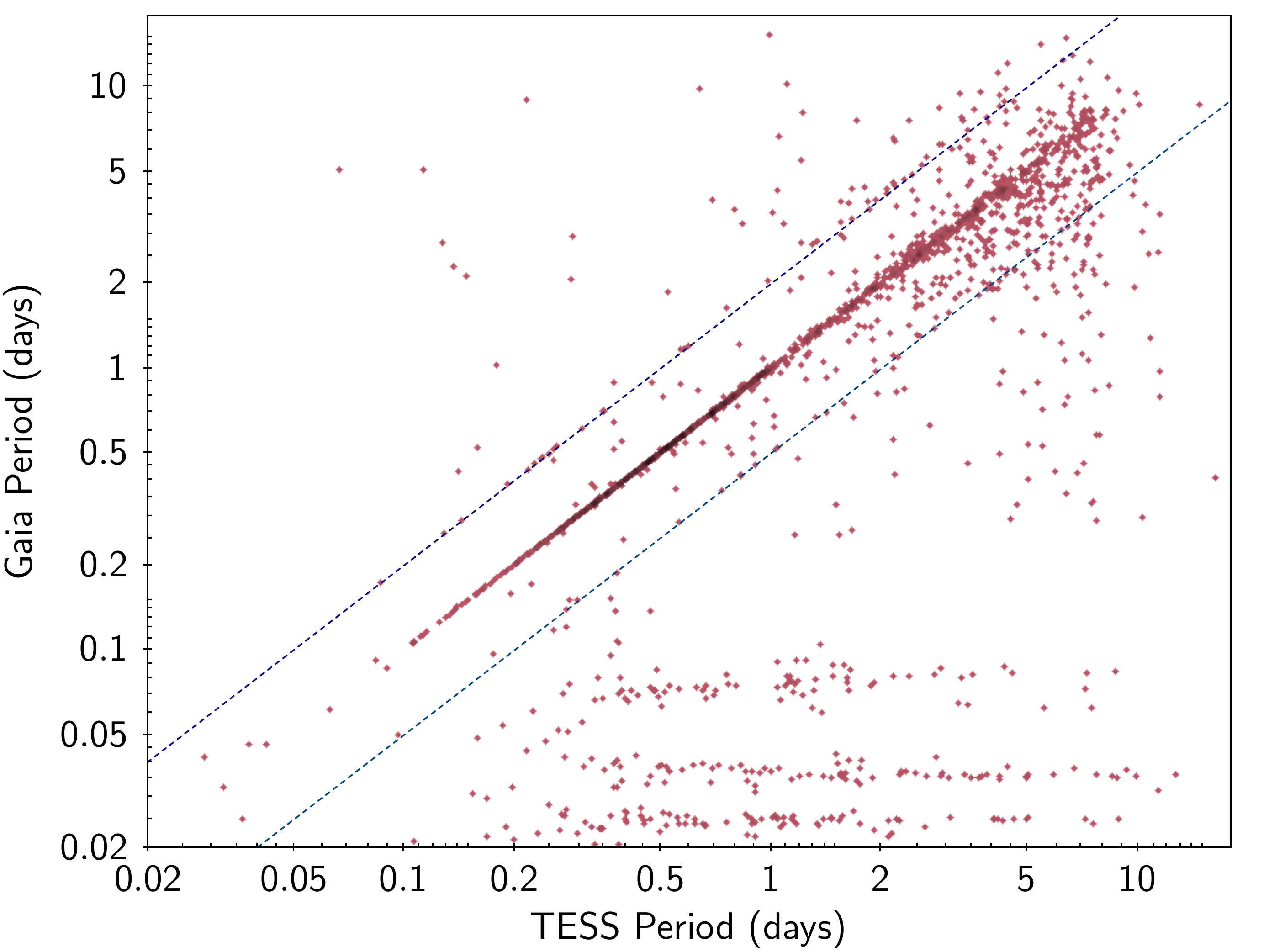}
\caption{Top left: comparison of periods presented in this work for the same sources observed in multiple sectors. Top right: comparison of periods in this work relative to periods in ASAS-SN survey \citep{jayasinghe2019}. Note the difference in beat frequency of 0.02 days in TESS and 1 day in ASAS-SN. Bottom left: comparison of periods to \citet{mcquillan2014} sample for the Kepler field. Dashed lines correspond to 2x and 0.5x harmonics. Bottom right: comparison of periods to Gaia DR3 periods \citep{eyer2022} from Short Timescale Variability and Rotational Modulation catalogs.
\label{fig:pairmatch}}
\end{figure*}

Approximately half of all sources have been observed in multiple sectors---most of which are due to repeated observations of Year 1 in Year 3. About 23\% of the stars in the sample are detected in more than two sectors, and $\sim$1\% are found in the TESS continuous viewing zone. As all of the sectors are treated independently, this provides an opportunity to check the internal quality of the measured periods from one sector to the next (see Figure~\ref{fig:pairmatch}a).

Performing an internal cross-match, we find that 89\% of sources have self-consistent periods between any two pairs of light curves when both had a period recovered. 7\% of pairs have harmonic period discrepancy by a factor of 2 (which is a common issue when searching for periodic signal), or they have a beat frequency of 0.02 days between the true period of a star with the minimum sampling of TESS FFIs (only affecting sources with periods $<$0.1 days in these data). Approximately 2\% of sources have one of the epochs in the pair dominated by the artificial period of $\sim$7 days, corresponding to the half an orbit for TESS. The remaining 2\% have periods that are uncorrelated between epochs. Thus, although there are sources that do have inconsistent periods, the vast majority of them ($>$95\%) do appear to be well-behaved (i.e., providing periods that agree directly, or with well understood harmonics).

However, of $\sim$150,000 pairs where periodicity was detected in at least one of the sectors, only 50\% of the sources have recovered periodicity in both sectors. Examining the light curves in the sectors for which the periodicity was not recovered shows that these stars usually continue to exhibit periodicity, but the recovery failed in most cases because of the noise. Different sectors of TESS data can have different noise properties due to the different orientations of the spacecraft, Moon, and Earth. In addition, the position of any given star on the detectors changes between sectors due to the rotation of the spacecraft. This rotation of the focal plane can yield different ``optimal apertures'' between sectors. The light curves in which periods were not recovered often have a greater degree of scatter, thus they fail to meet the strict criteria outlined in Section~\ref{sec:periods}. An additional astrophysical effect that might be relevant is that the intrinsic amplitude of starspot-induced variability does evolve over time, in concert with the spot-covering fractions on the stellar surface \citep{basri2020}.

\begin{figure}[!ht]
\includegraphics[width=0.9\linewidth]{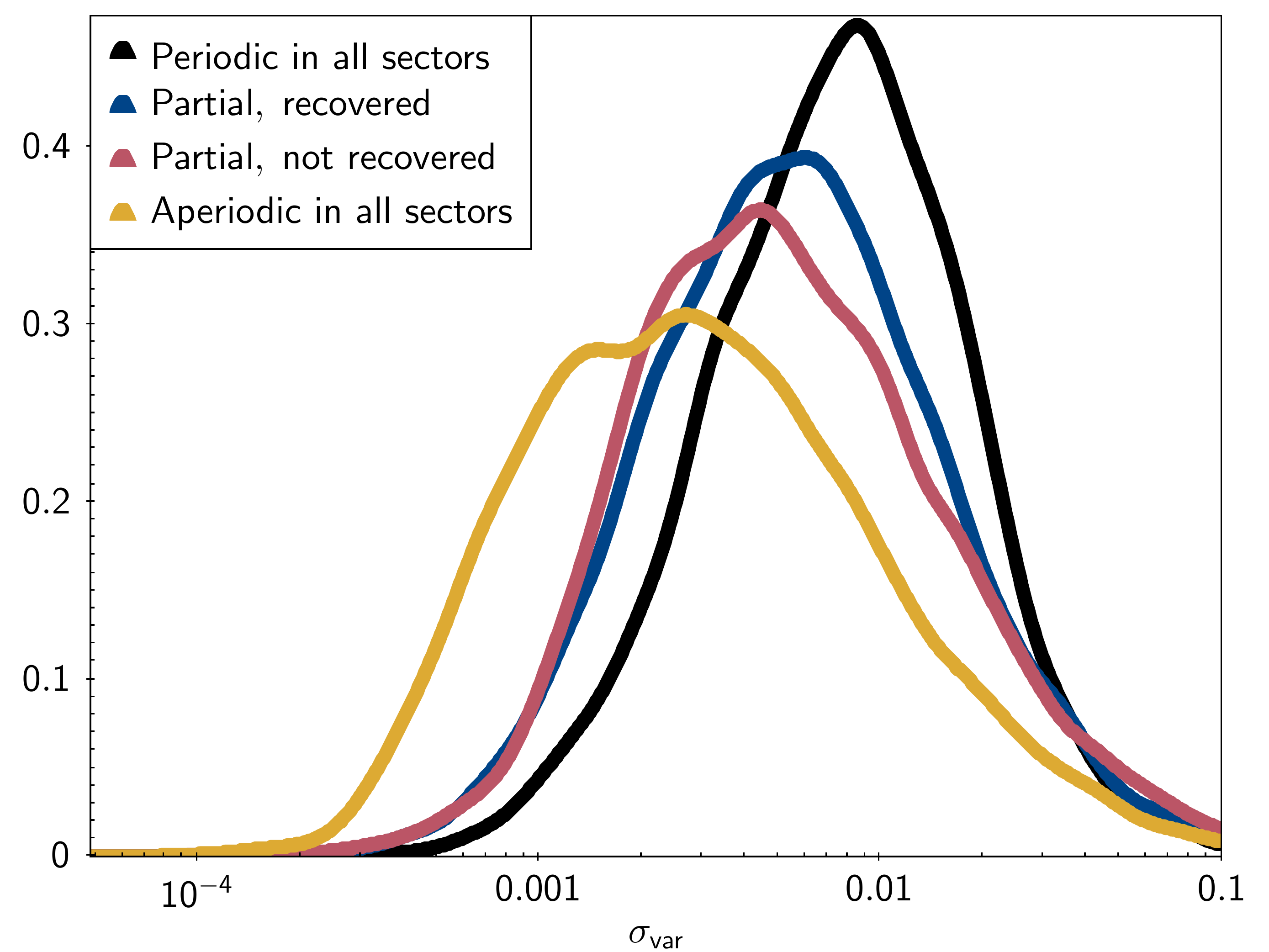}
\includegraphics[width=0.9\linewidth]{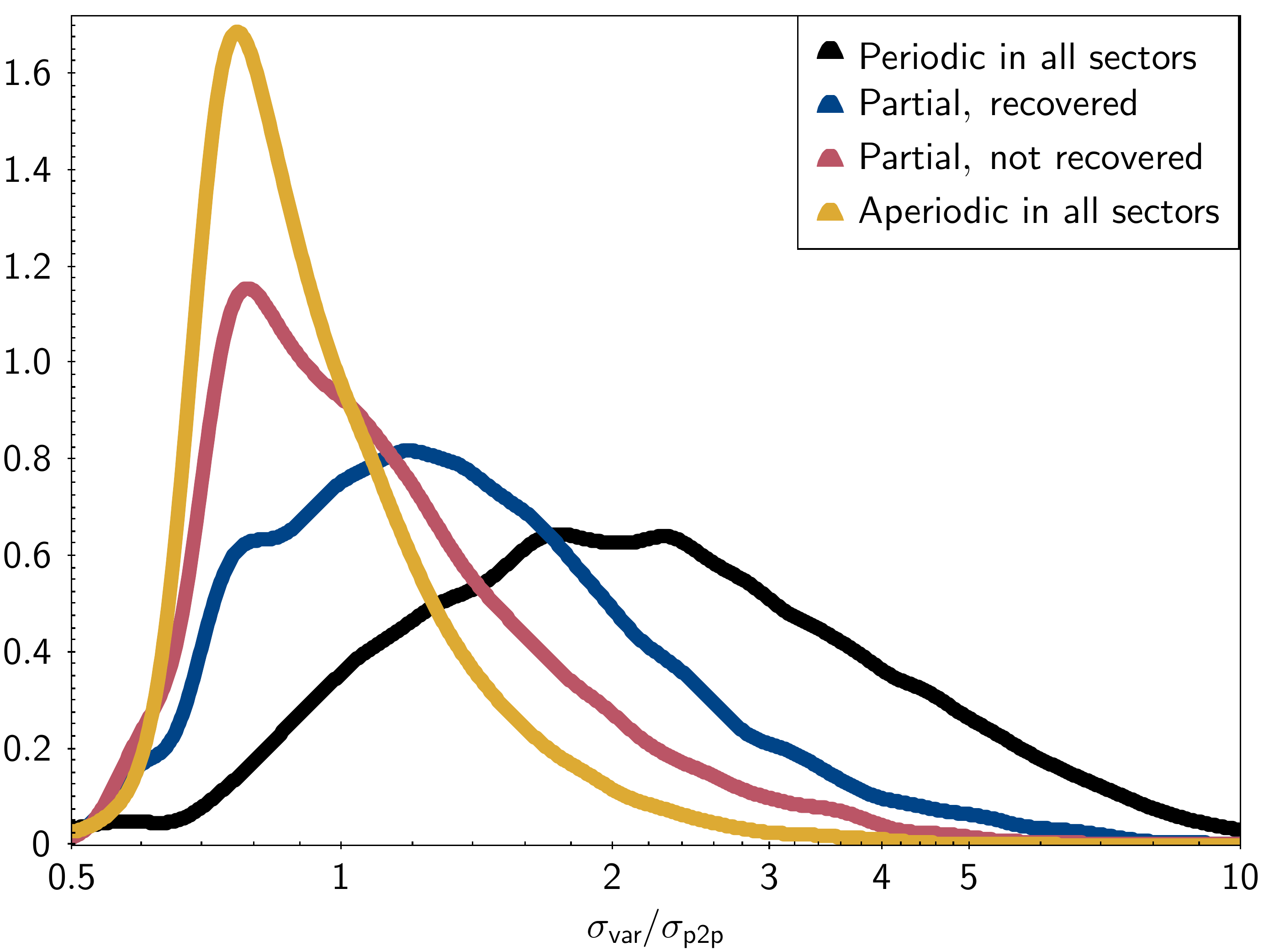}
\includegraphics[width=0.9\linewidth]{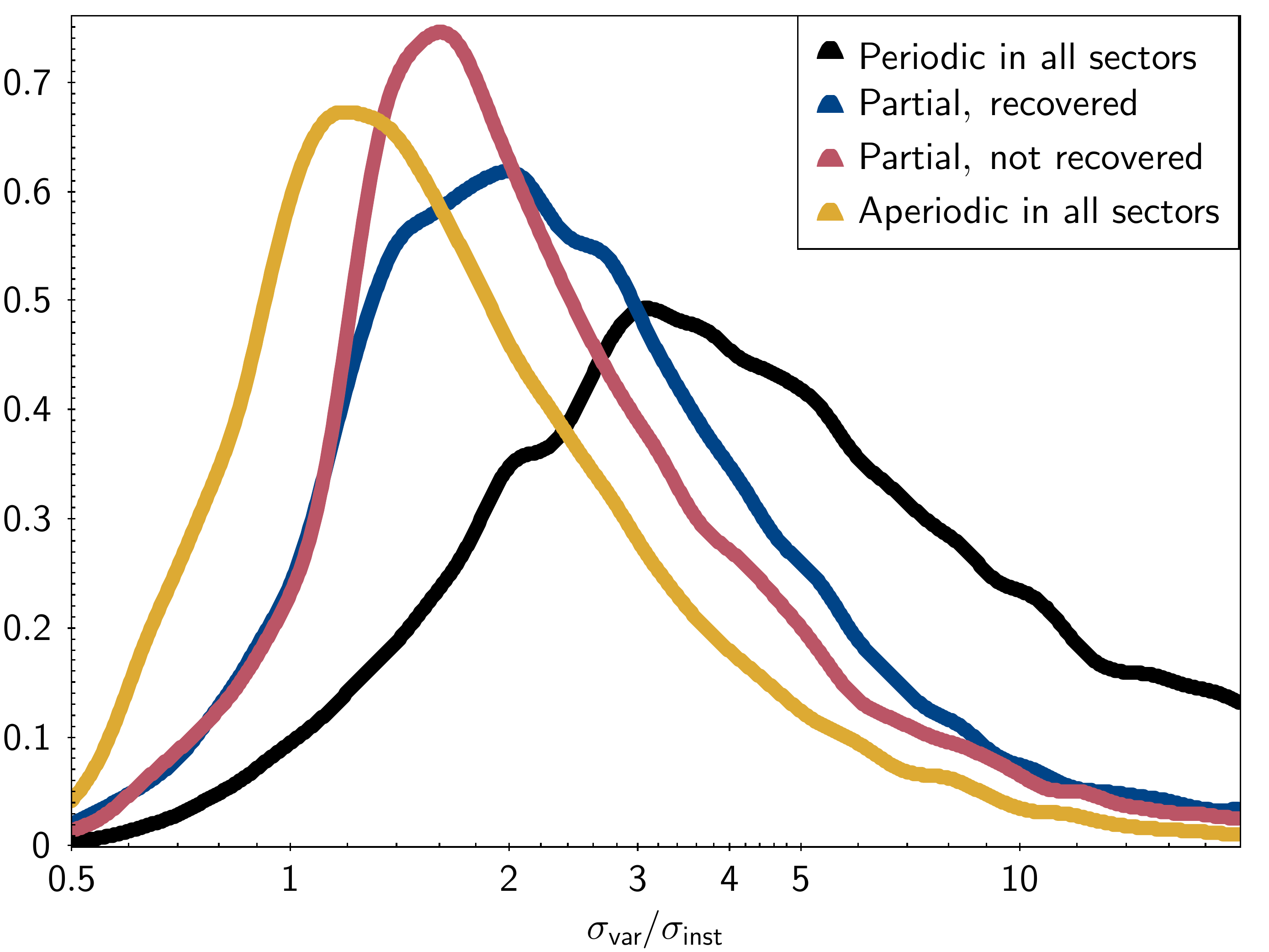}
\caption{Distributions of the light curve standard deviations ($\sigma_{\rm var}$), standard deviation normalized by the point-to-point RMS ($\sigma_{\rm var}/\sigma_{\rm p2p}$), and standard deviation normalized by the reported mean flux uncertainty ($\sigma_{\rm var}/\sigma_{\rm inst}$). Sources for which periodicity was recovered in all sectors are shown in black, and sources that appear to be aperiodic in all sectors are in yellow. Sources that were detected as periodic in some but not all sectors are shown in blue and red, with blue corresponding to the specific sectors in which the source is recovered as periodic, and red corresponding to the sectors where the same sources are not recovered as periodic.
\label{fig:sigmavar}}
\end{figure}

For each light curve we record the typical $\sigma_{\rm inst}$, which corresponds to the typical uncertainty reported by \texttt{eleanor} normalized relative to the flux. We also measure $\sigma_{\rm p2p}$ corresponding to the point-to-point scatter in the normalized light curve that is not affected by the amplitude imposed by large scale (periodic) variability. We then compare it to $\sigma_{\rm var}$ which is a standard deviation of the light curve. We find that the sources in which periodicity was recovered $\sigma_{\rm var}/\sigma_{\rm p2p}$ tends to be significantly higher than in the sources with partial period recovery in only one of the sectors. The sectors in which period recovery for a given source has occurred have higher $\sigma_{\rm var}/\sigma_{\rm p2p}$ than in the sectors where it did not, however, they tend to have a more comparable $\sigma_{\rm var}$, as well as $\sigma_{\rm inst}$ (Figure \ref{fig:sigmavar}).

This is driven by two factors. First, higher amplitude of variability $\sigma_{\rm var}$ makes it more robust against both noise and intrinsic amplitude evolution. Second, period recovery is more likely to occur in sources with higher SNR. The higher mass stars tend to be recovered as periodic more reliably than lower mass stars (as such, excluding periodic sources with, e.g., $\sigma_{\rm var}/\sigma_{\rm p2p}>1$ or $\sigma_{\rm var}/\sigma_{\rm inst}>2$ would only preferentially exclude lower mass stars). And, similarly, a light curve with higher SNR in one sector for the same star is more likely to be recovered to be periodic.

We compare the recovered periods to periods measured by ASAS-SN \citep{jayasinghe2019}, see Figure~\ref{fig:pairmatch}b. There are 3,766 pairs of observations between periods presented here, and the sources in ASAS-SN. Of these, 90\% are classified as rotational variables or as YSOs. 8\% are classified as eclipsing binaries (3\% detatched, 2\% semi-detatched, 4\% contact), and 1\% are pulsating variables.

Periods between ASAS-SN and TESS tend to be in good agreement, with 74\% of sources having identical period, 15\% having a harmonic period offset by a factor of 2. Five percent of sources have a a beat frequency of $\sim$1 day, driven by the cadence in ASAS-SN observations: unlike the beat frequency of TESS, it can affect periods up to 10 days \citep[for an example of a similar phenomenon see also ][]{covey2016}.

Similarly, we compare periods to those measured in the Kepler field \citep{mcquillan2014}, see Figure~\ref{fig:pairmatch}c. Eighty percent of 108 sources in common have good agreement. Unlike the internal match, or a match against ASAS-SN, almost no sources have a harmonic period offset, with an exception of those sources with periods $>$10 days in Kepler, which get aliased to the 2:1 harmonic.

Finally, we compare the recovered rotational periods in this work to the recently released periods from Gaia DR3 \citep{eyer2022}. Gaia has a relatively sparse light curve (often containing as few as 30 measurements). Given such sparsity of measurements, the quality of the recovered periods by Gaia can be surprising, but there is a good agreement between them and the periods measured in this work with TESS for sources with periods $>0.1$ days. At the periods reported by Gaia $<$0.1 days, there is little to no agreement, in almost all cases there is a much more pronounced longer period signal that is recovered in TESS. 

\begin{figure}[!t]
\includegraphics[width=\linewidth]{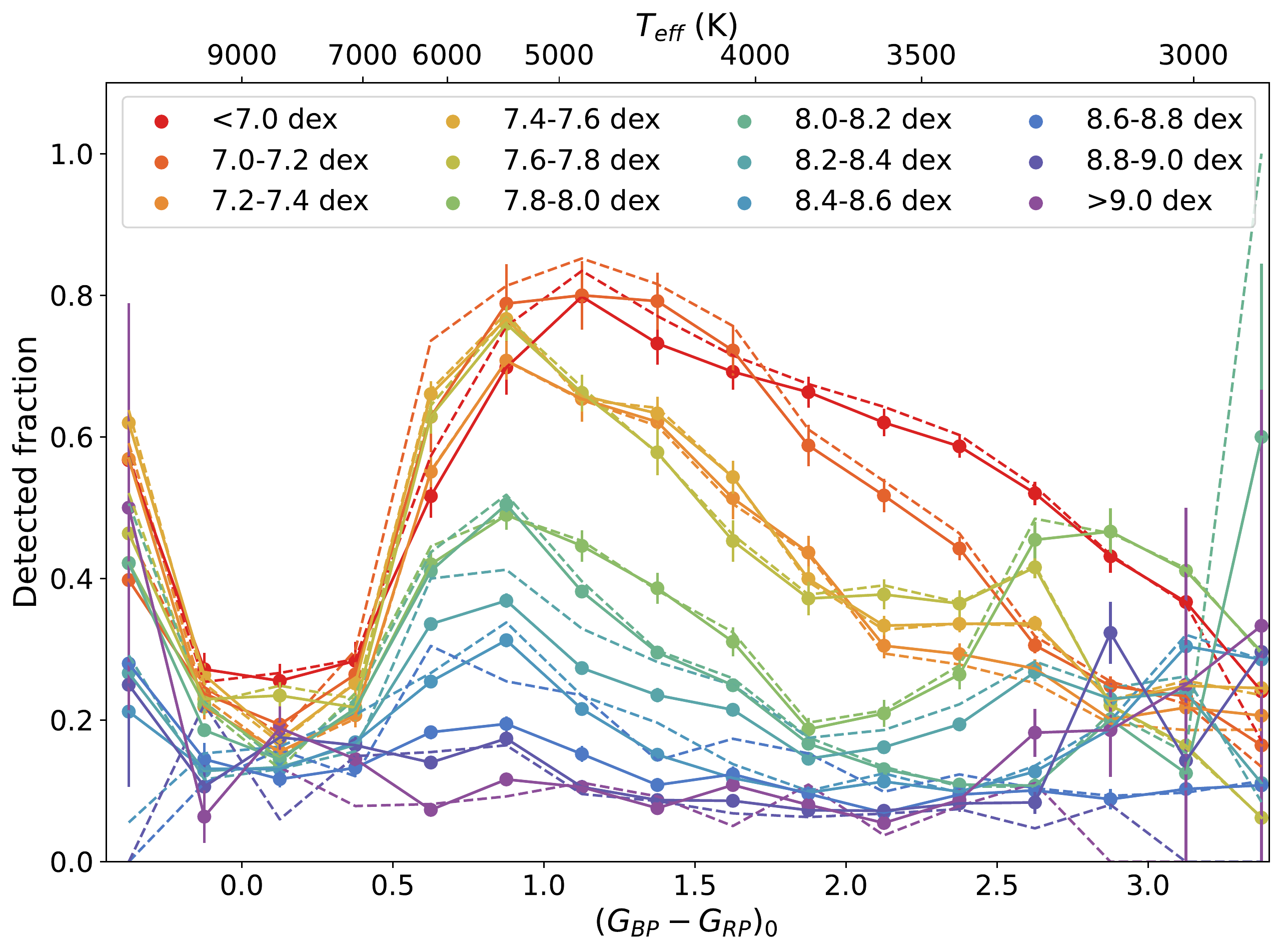}
\caption{Fraction of sources as a function of Gaia color and age in which confident periodic signal (up to 12 days) is observed in TESS data. The dashed line shows the sample solely among the strings of stars, defined in \citet{kounkel2019a}. Note that in almost all of the cases the stellar strings have members with similar fraction of the periodic variables as the full sample, and it is usually somewhat higher.
\label{fig:fraction}}
\end{figure}

\begin{deluxetable*}{ccl}[!ht]
\tablecaption{Measured TESS periods
\label{tab:data}}
\tabletypesize{\scriptsize}
\tablewidth{\linewidth}
\tablehead{
 \colhead{Column} &
 \colhead{Unit} &
 \colhead{Description}
 }
\startdata
Gaia id & & Gaia DR2 source ID \\
TIC & & TESS input catalog ID \\
RA & deg & Right ascention in J2000 \\
Dec & deg & Declination in J2000 \\
Sector & & TESS sector \\
Theia ID & & Group id from \citet{kounkel2020} \\
Age & log yr & Average age of the stellar group \\
e\_age & log yr & Uncertainty in age \\
c\_bp\_rp & mag & Extinciton corrected Gaia BP-RP color \\
c\_absg & mag & Extinction corrected Gaia G magnitude \\
Tmag & mag & TESS magnitude \\
Teff & K & Effective temperature from TESS Input Catalog \\
e\_teff & K & Uncertainty in Teff \\
Mass & Msun & Mass from TESS Input Catalog \\
e\_mass & Msun & Uncertainty in mass \\
Radius & Rsun & Radius from TESS Input Catalog \\
e\_radius & Rsun & Uncertainty in radius \\
corr\_p\_1 & day & Primary period in CORR light curves \\
e\_corr\_p\_1 & day & Uncertainty in corr\_p\_1 \\
corr\_p\_2 & day & Secondary period in CORR light curves \\
e\_corr\_p\_2 & day & Uncertainty in corr\_p\_2 \\
corr\_a\_1 & & Primary amplitude in CORR light curves \\
corr\_a\_2 & & Secondary amplitude in CORR light curves \\
pca\_p\_1 & day & Primary period in PCA light curves \\
e\_pca\_p\_1 & day & Uncertainty in pca\_p\_1 \\
pca\_p\_2 & day & Secondary period in PCA light curves \\
e\_pca\_p\_2 & day & Uncertainty in pca\_p\_2 \\
pca\_a\_1 & & Primary amplitude in PCA light curves \\
pca\_a\_2 & & Secondary amplitude in PCA light curves \\
raw\_p\_1 & day & Primary period in RAW light curves \\
e\_raw\_p\_1 & day & Uncertainty in raw\_p\_1 \\
raw\_p\_2 & day & Secondary period in RAW light curves \\
e\_raw\_p\_2 & day & Uncertainty in raw\_p\_2 \\
raw\_a\_1 & & Primary amplitude in RAW light curves \\
raw\_a\_2 & & Secondary amplitude in RAW light curves \\
Period & day & Adopted period from the light curve \\
e\_period & day & Uncertainty in period \\
ptype & & Adopted period source \\
L & log kg m$^2$ s$^{-1}$ & Calculated angular momentum \\
e\_L & log kg m$^2$ s$^{-1}$ & Uncertainty in angular momentum \\
RUWE & & Renormalised unit weight error from Gaia EDR3 \\
sigma\_var & & Standard deviation of the normalized flux (full amplitude) \\
sigma\_inst & & Typical reported uncertainty in the flux of the normalized flux (instrumental) \\
sigma\_p2p & & 68th percentile point-to-point variability of the normalized flux (measured scatter) \\
c1 & & Flux contamination ratio within 10$''$ \\
c2 & & Flux contamination ratio within 20$''$ \\
c3 & & Flux contamination ratio within 30$''$ \\
c4 & & Flux contamination ratio within 40$''$ \\
c5 & & Flux contamination ratio within 50$''$ \\
c6 & & Flux contamination ratio within 60$''$
\enddata
\end{deluxetable*}

Figure~\ref{fig:fraction} presents an overview of the rate of period detection as a function of stellar color (optical, Gaia passbands) and age (inherited from the moving group membership, using the isochrone fitting of the photometry of all the stars in the group). There is a clear dependence of the recovery of the periods on age, with youngest stars being most variable, and mass, with solar-type stars being most variable. The decrease towards lower mass stars may be partially attributable to the lower signal-to-noise due to the increased faintness. The decrease of periodicity with mass is not monotonic for stars older than 7.8 dex, developing a secondary local maxima at colors corresponding to M dwarfs. Periodicity also sharply decreases towards early type stars as the dominant mode of variability switches from rotation to pulsation.

Finally, we compare the frequency of periodicity recovery in stellar strings, the extended comoving groups that were identified in \citetalias{kounkel2019a}. These strings are ubiquitous at ages $<100$ Myr as they consist of populations that still retain their primordial morphology that have not yet been fully dissolved in the field. We find that the stars found in strings tend to have either comparable or greater fraction of variable stars compared to the full sample. Thus, they tend to have a greater degree of variability compared to the more compact, isolated, and cluster-like groups lacking diffuse halos, even after accounting for the age/mass dependency. This could perhaps be due to a few factors---(i) the environment could impact the rotation period distributions to a considerable degree for stars of the same age \citep[however, see][]{Fritzewski2020}, (ii) the TESS light curves for stars in denser clusters are more strongly affected by source confusion and signal interference, (iii) or the sample of strings have a smaller degree of contamination from field stars than the more compact groups. We note that the level of contamination is expected to be on a few percent level, and that it is somewhat more pronounced at the edges of the diffuse halos in comparison to the kinematic ``core'' of a given population \citep{bouma2021}.

We further discuss the the physical significance of the trends in ages and masses in the next section.

\section{Results}\label{sec:results}

\begin{figure*}
\epsscale{1.0}
\plotone{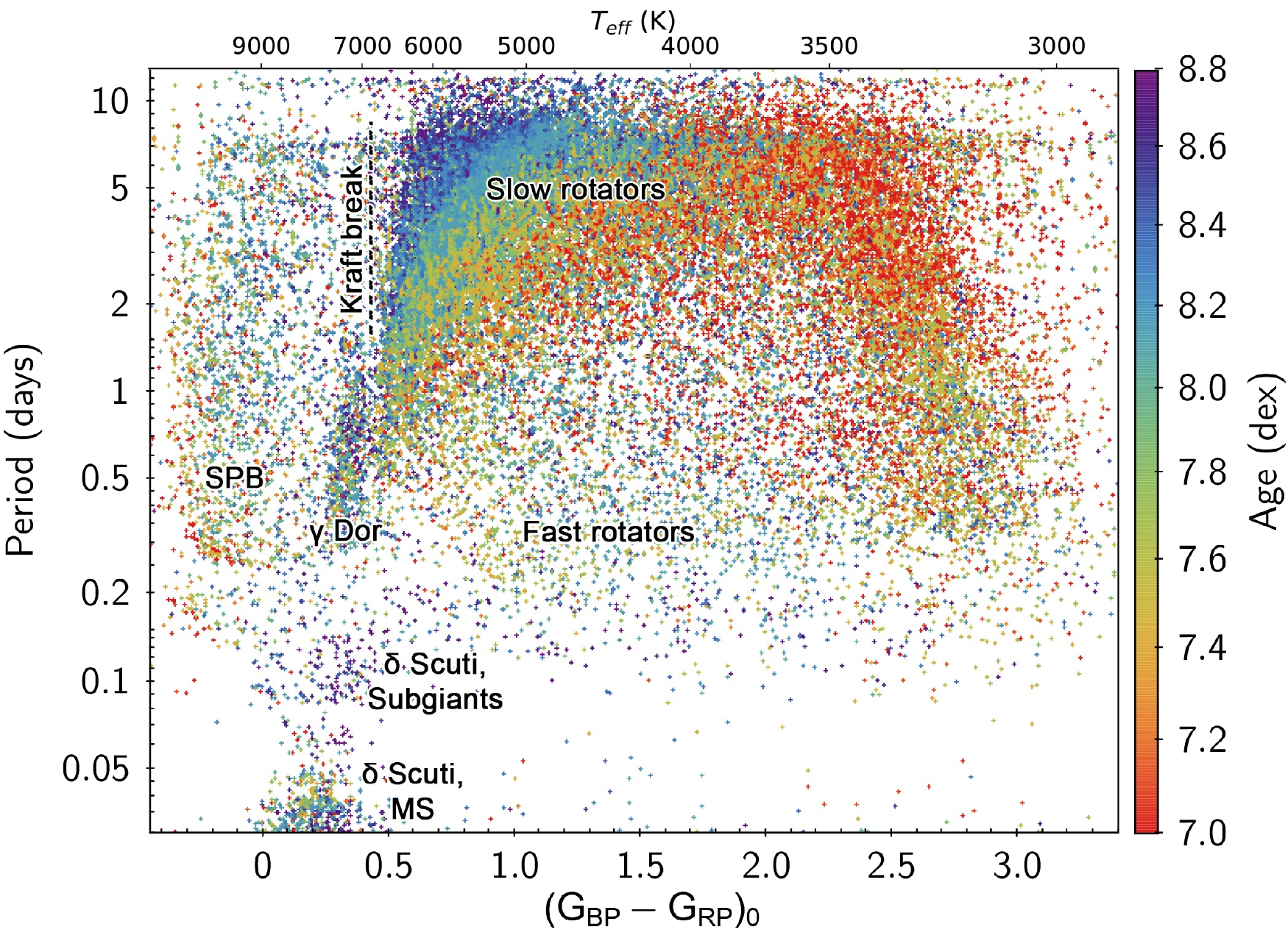}
\caption{Recovered distribution of periods of the stars in the sample as a function of Gaia color, extinction corrected, color-coded by their age, annotated with the features of the parameter space. Sources redder than the labelled ``Kraft break'' have the dominant mode of variability being caused by rotation. They can be separated into slow rotators (sources that show a clear evolution of their periods with age), and fast rotators, having typical periods $<$1 day. Sources bluer than the Kraft break are commonly pulsators. $\gamma$ Dor sources have pulsation frequency comparable to the stellar rotation period, consisting only of the main sequence stars. $\delta$ Scuti stars in the age limited sample form a bimodal distribution, with the main sequence $\delta$ Scuti variables having periods shorter than 1.2 hours, and the subgiants having periods longer than 2 hours. Slowly pulsating B-type variables (SPB) are found among the early type stars. 
\label{fig:periodogram}}
\end{figure*}

Figure~\ref{fig:periodogram} presents the fundamental result of this work: the empirical distribution of periods (determined from TESS light curves; see Sections~\ref{sec:tess} and \ref{sec:periods}) as a function of stellar color and age (determined from the Theia catalog; see Section~\ref{sec:theia}). All the colors have been corrected by the estimated extinction in \citet{kounkel2020}. At $(G_{\rm BP} - G_{\rm RP})$ color of 0.4, there appears to be a gap in the distribution of periodic variables. This gap most likely corresponds to the well-known Kraft break \citep{kraft1967,avallone2022}, separating the stars with convective and radiative envelopes. One surprise however is that we see the gap at $\approx$6700 K, whereas the Kraft break is traditionally placed at $\approx$6300 K, based on a sharp transition in stellar equatorial velocity as a function of mass or temperature. While we believe the same underlying physics to be at play, the general implication is likely that stars do in fact remain spotted up to mildly larger masses ($\approx$1.4\,\msun) than the traditional $\approx$1.25\,\msun cutoff.

Our primary purpose is to understand the evolution of stellar angular momentum and the refinement of relations for gyrochronology. Thus, for the remainder of our analysis we restrict our discussion to stars with convective envelopes, which experience wind-driven rotational evolution and for which gyrochronology can be applied. An overview of the periodic variables with radiative envelopes is provided in Appendix~\ref{sec:radiative}. 
The full sample of stars with rotation periods and ages resulting from our analysis is presented in Table~\ref{tab:data}.


\subsection{A catalog of rotation periods and ages in the field}

At the most basic level, for stars with $(G_{\rm BP} - G_{\rm RP})>$ 0.4, Figure~\ref{fig:periodogram} manifests a clear morphology and gradient that represents (a) the dependence of stellar rotation period with stellar color (i.e., stellar temperature or mass), and (b) the evolution of this mass--rotation relationship with stellar age (i.e., stars across this mass range spin down over time). 
We note that all of the quantities in Figure~\ref{fig:periodogram} are empirically determined and are moreover determined independently of one another. The colors are measured directly, the periods are determined by us here from light curves obtained by TESS, and the ages have been previously estimated from a spatial-kinematic clustering analysis of Gaia data \citep{kounkel2020}. This represents the largest collection to date of stars in the field with empirically and independently determined rotation periods and ages. 

Because the spin-down of stars with age is well established observationally and theoretically, we can on the one hand regard the clear gradient in Figure~\ref{fig:periodogram} as a fundamental validation of the ages (or at least the relative time ordering) inferred by the Theia analysis of stellar associations. On the other hand, taking the Theia ages at face value, Figure~\ref{fig:periodogram} can be regarded as an affirmation of the soundness of gyrochronology as an approach to inferring stellar ages in the field. 

In the subsections that follow, we make use of this unique sample (Table~\ref{tab:data}) to revisit and refine empirical gyrochronology relations and to characterize the evolution of magnetic features on the stellar surfaces. 

\subsection{The slow sequence in the color--period diagram}

\begin{figure*}
\epsscale{1.0}
		\gridline{\fig{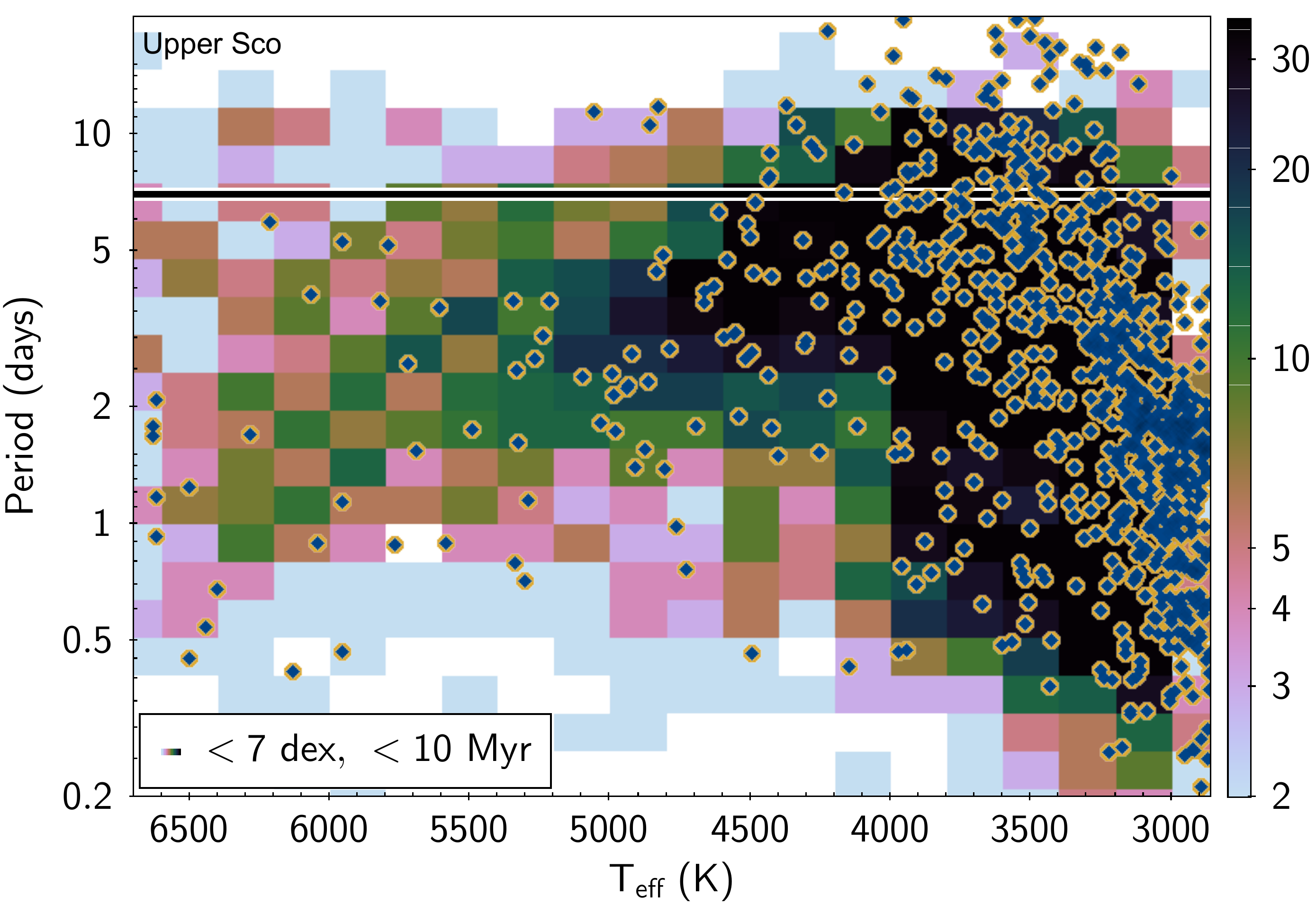}{0.33\textwidth}{}
		 \fig{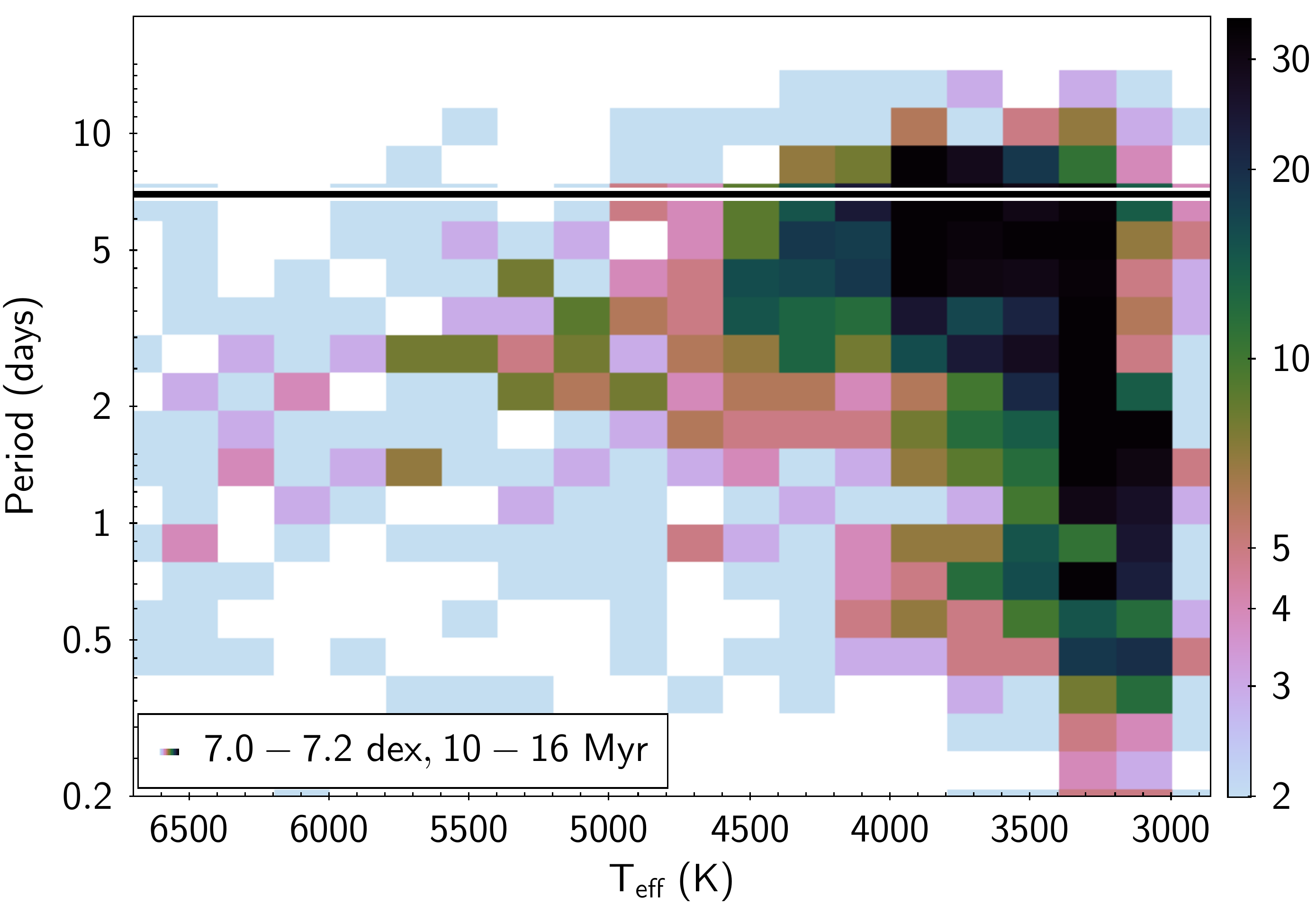}{0.33\textwidth}{}
		 \fig{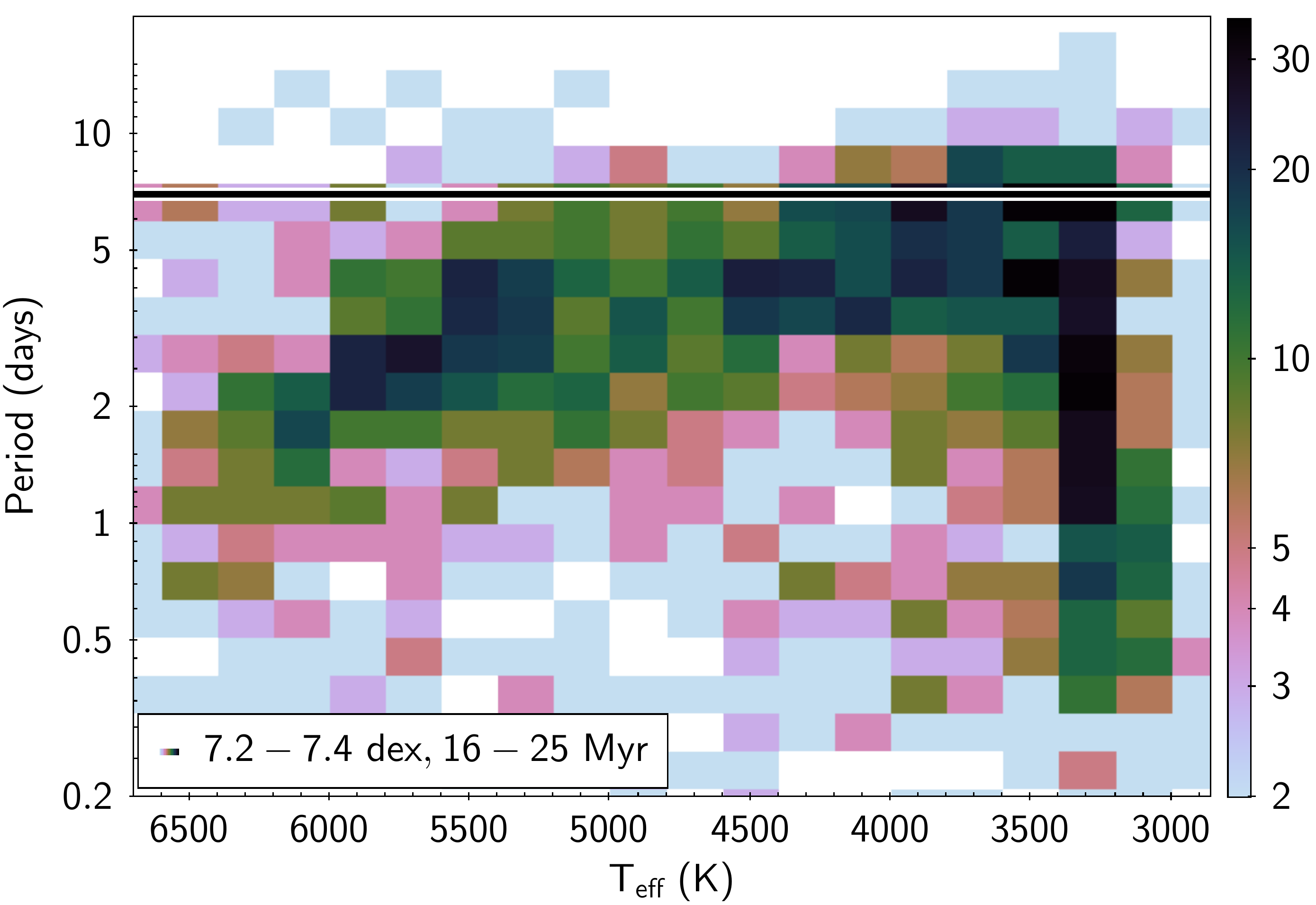}{0.33\textwidth}{}
 }\vspace{-0.5cm}
 \gridline{\fig{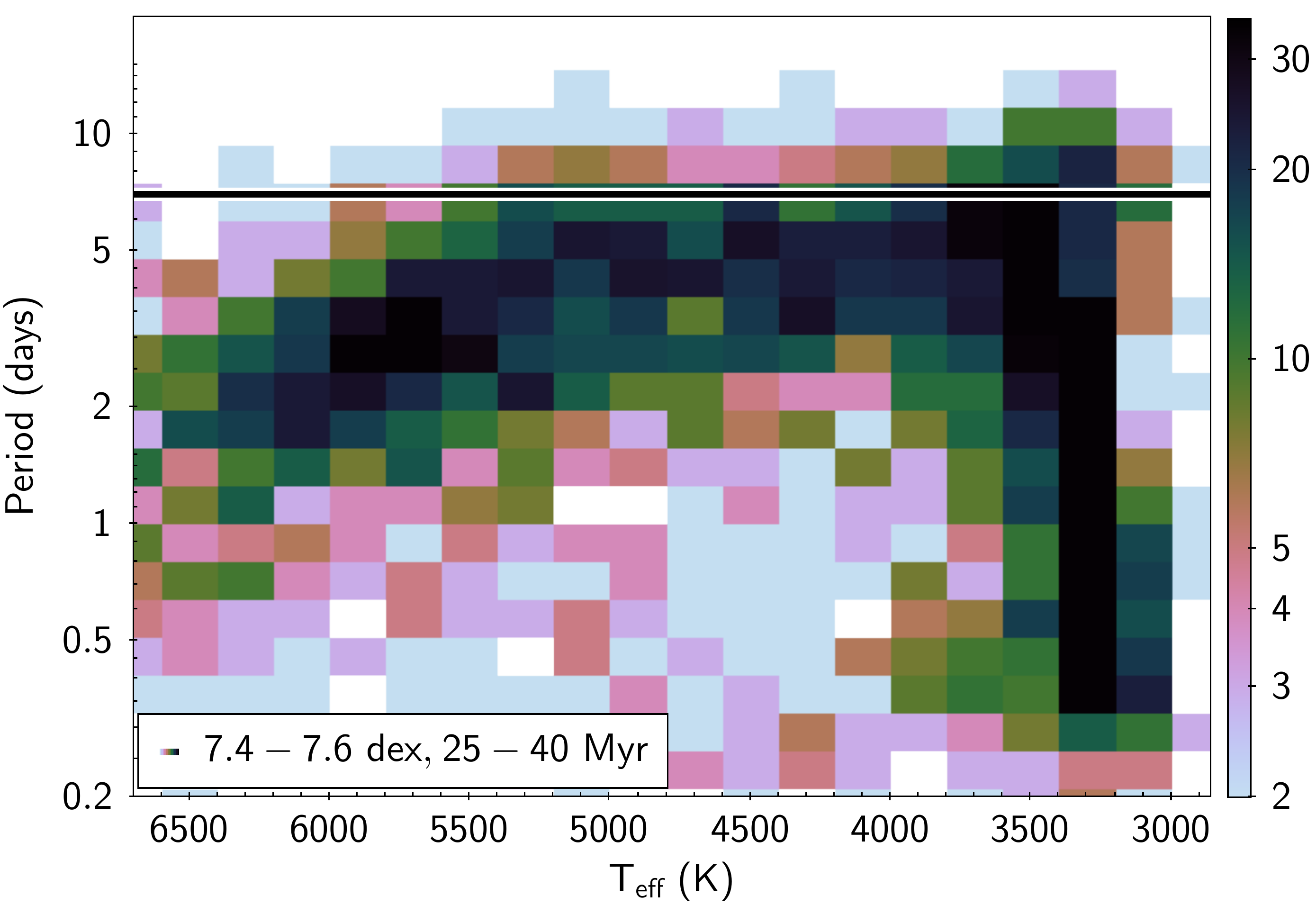}{0.33\textwidth}{}
		 \fig{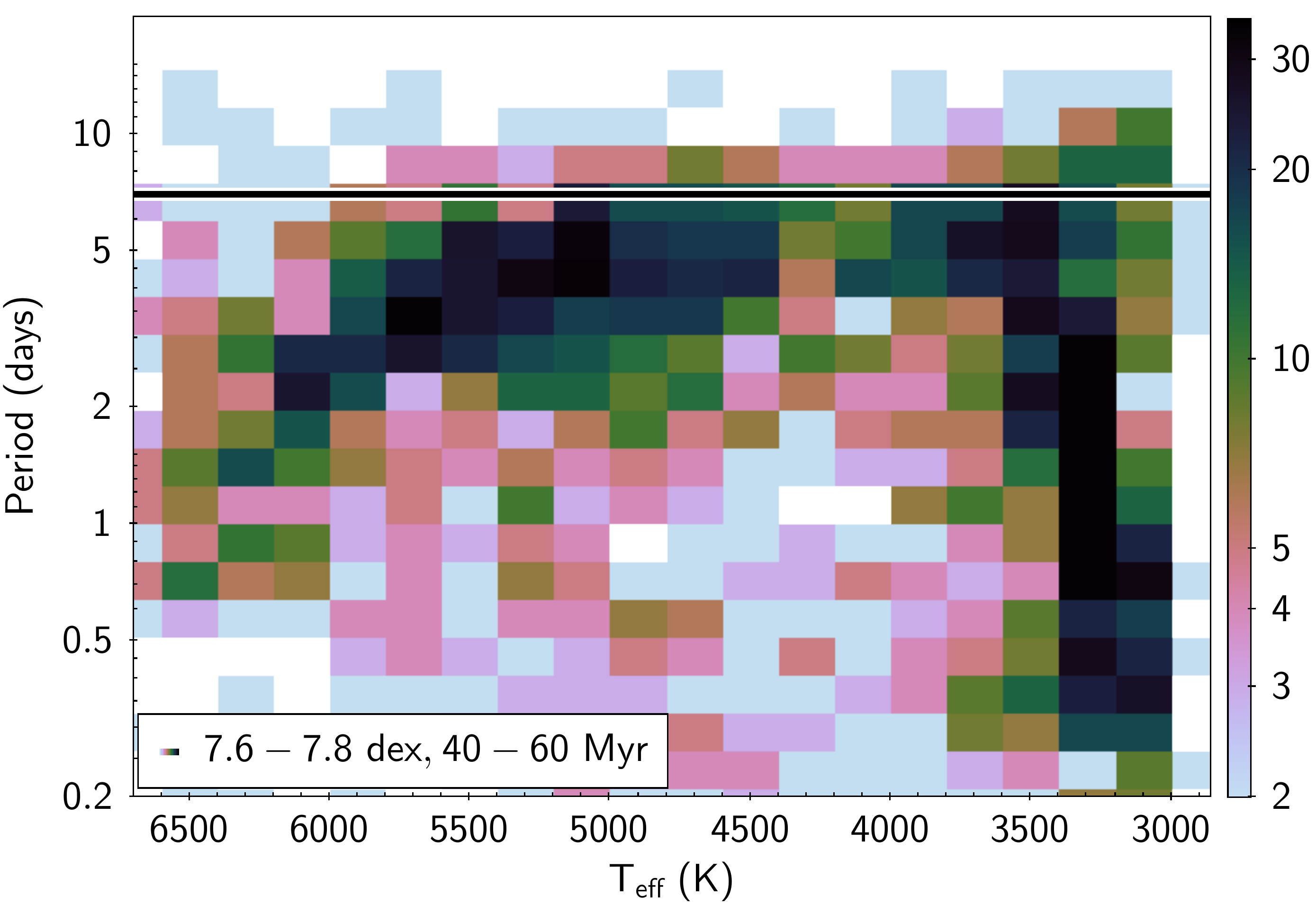}{0.33\textwidth}{}
		 \fig{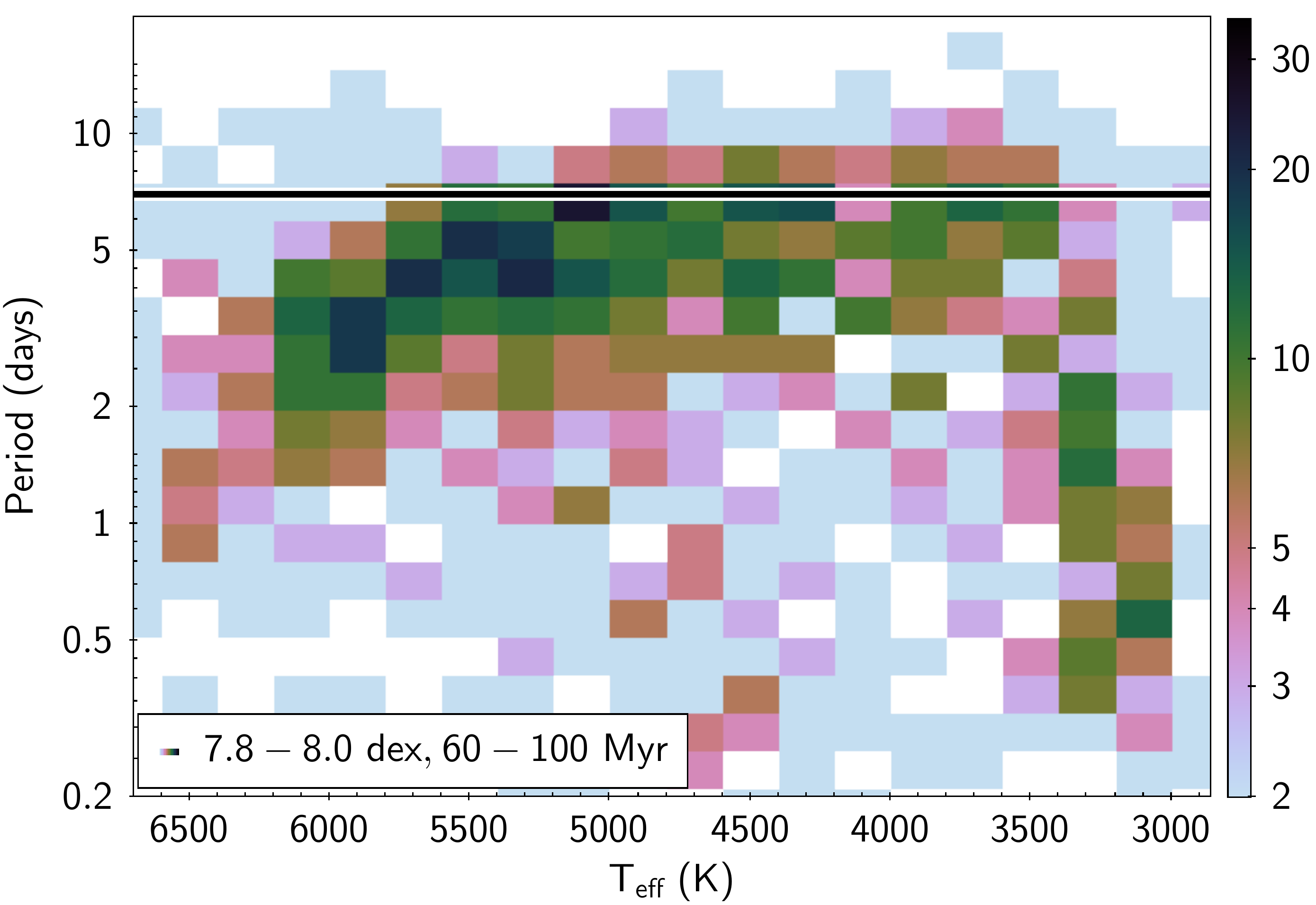}{0.33\textwidth}{}
 }\vspace{-0.5cm}
 \gridline{\fig{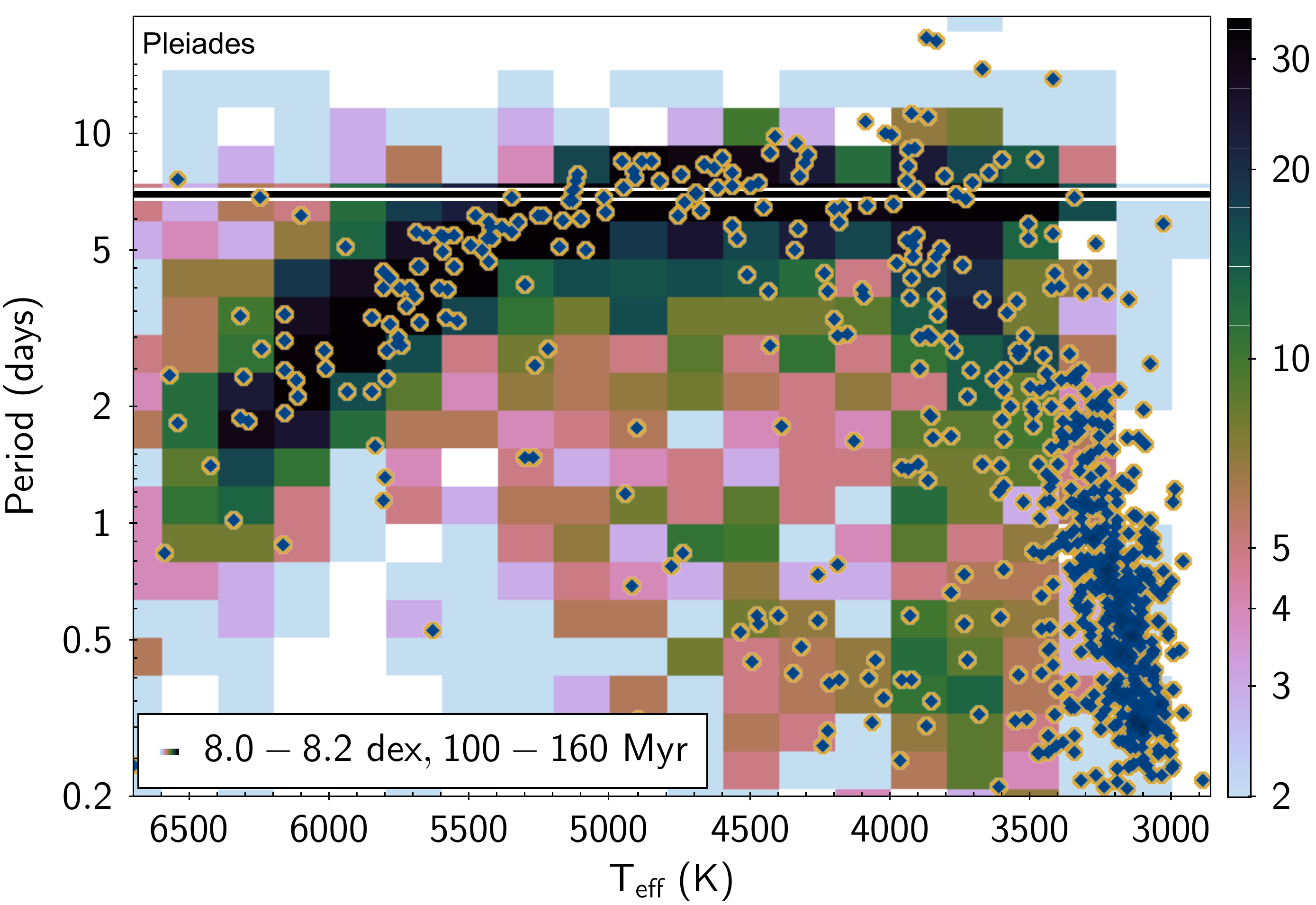}{0.33\textwidth}{}
		 \fig{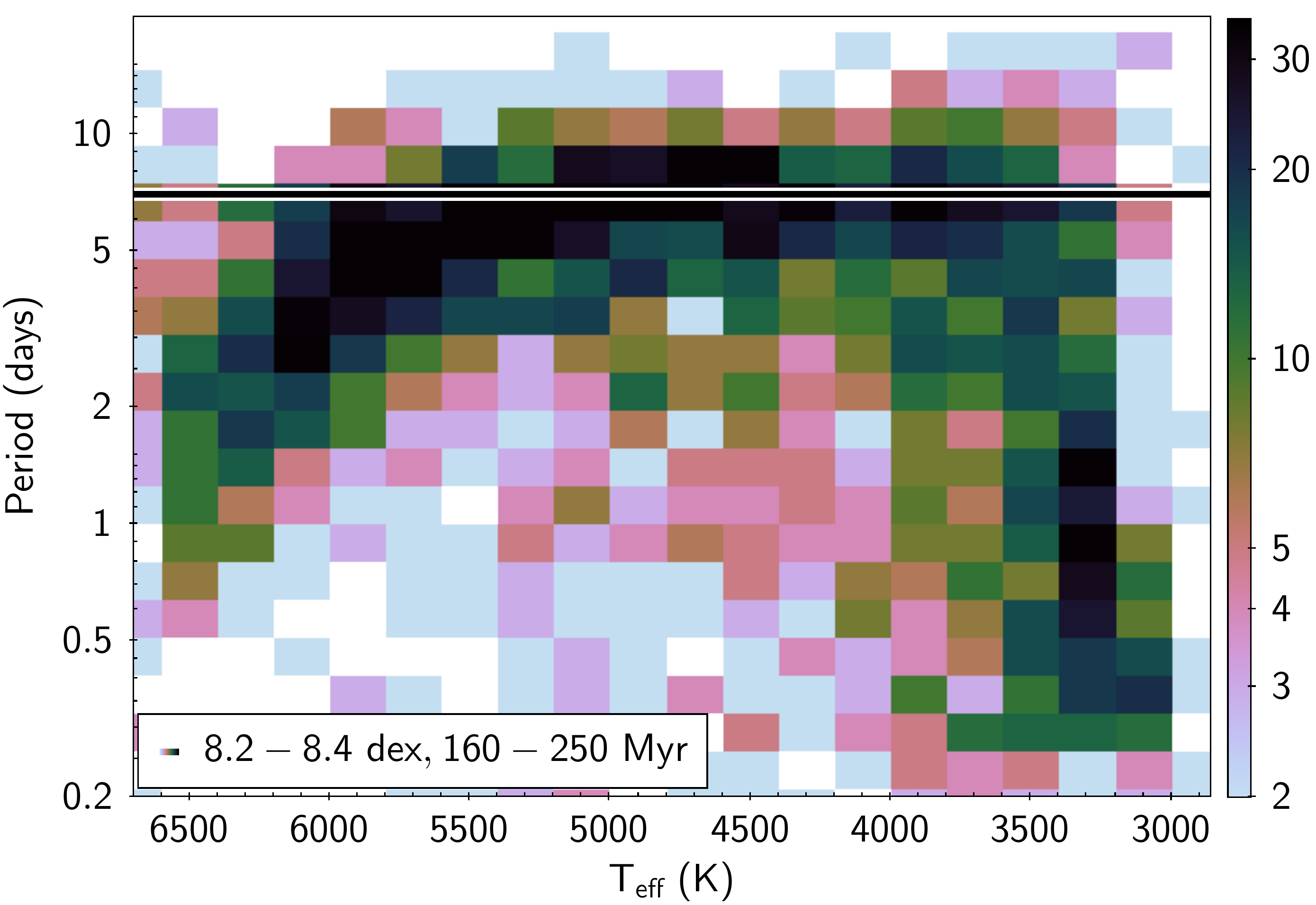}{0.33\textwidth}{}
		 \fig{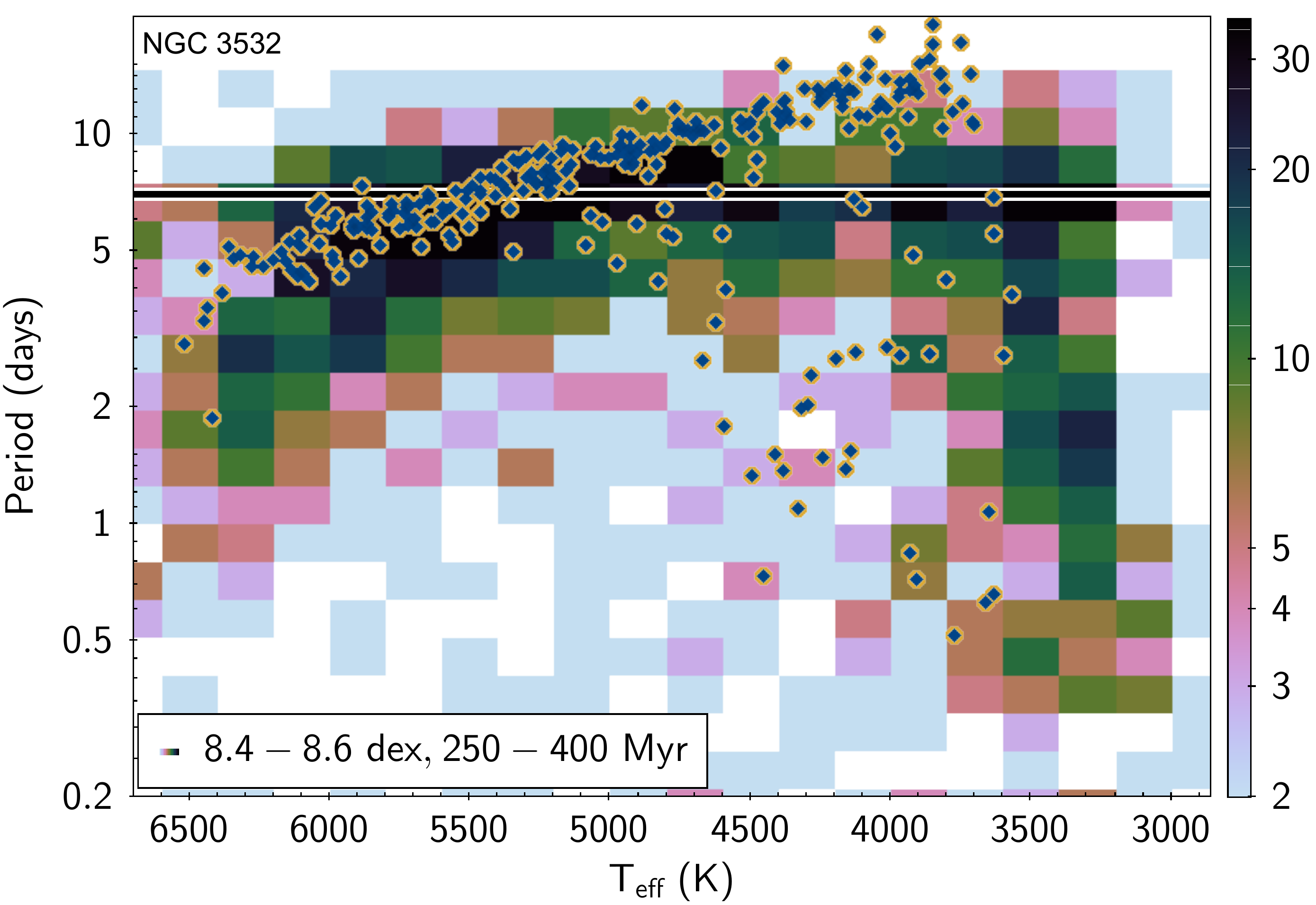}{0.33\textwidth}{}
 }\vspace{-0.5cm}
 \gridline{\fig{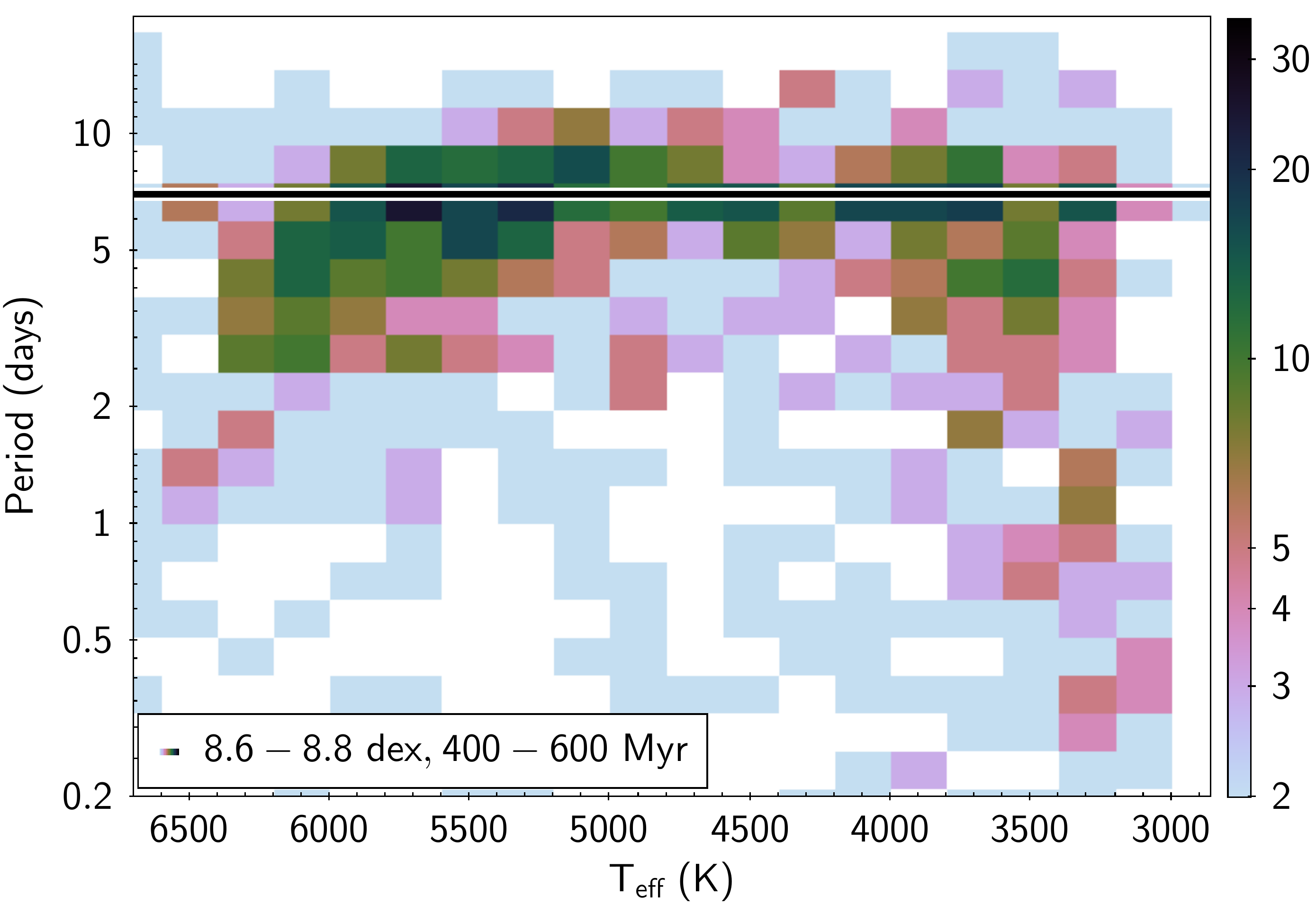}{0.33\textwidth}{}
		 \fig{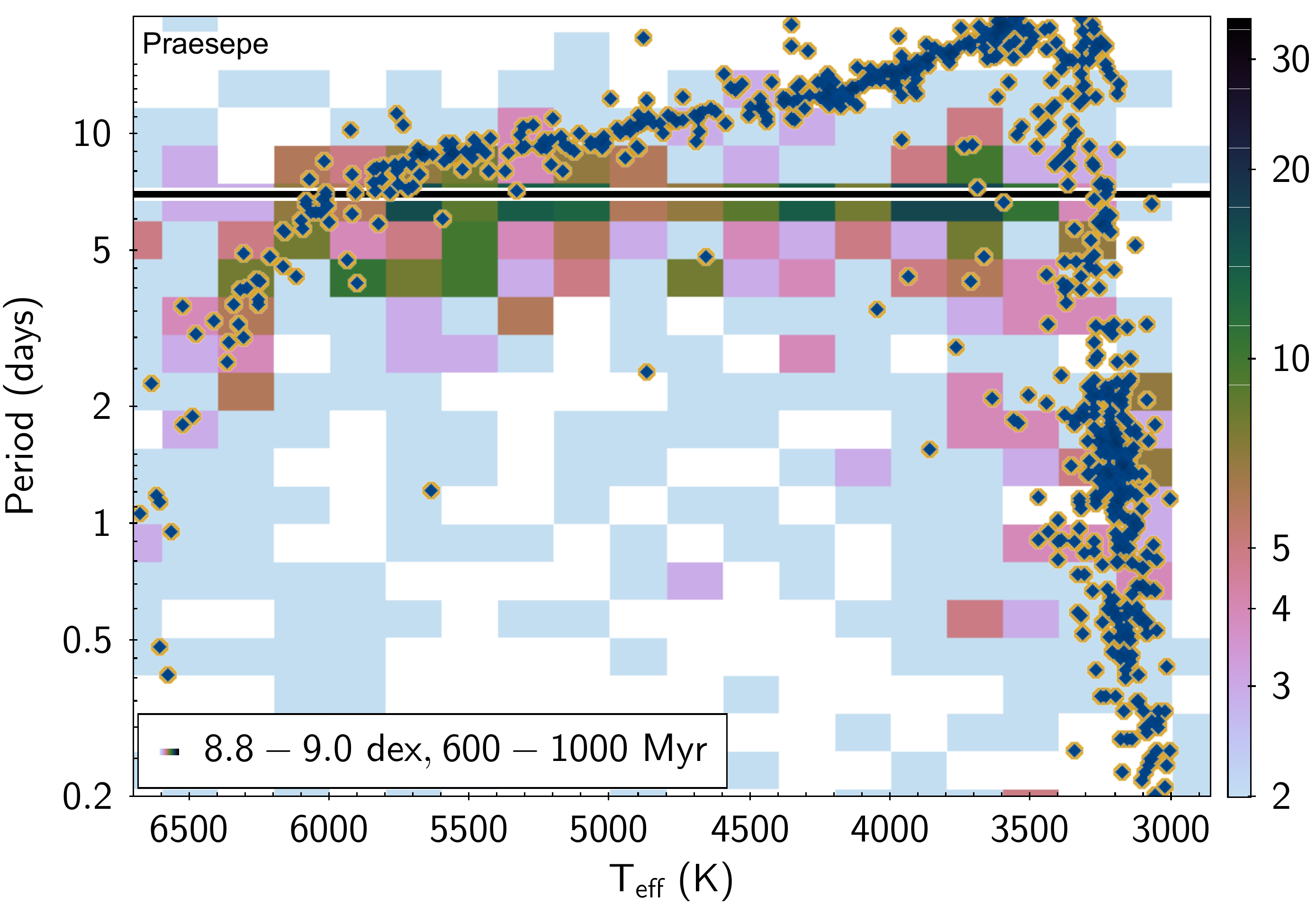}{0.33\textwidth}{}
		 \fig{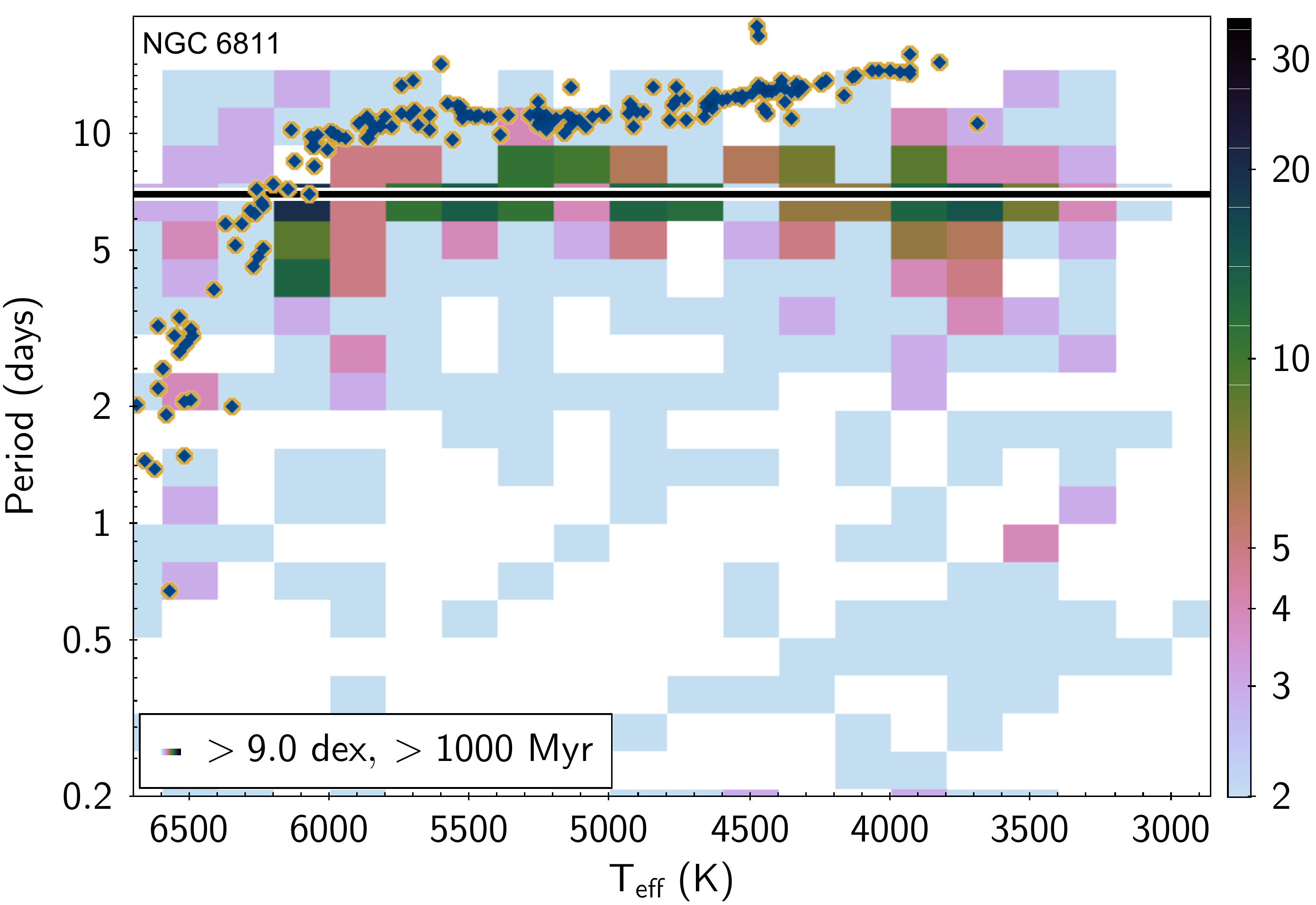}{0.33\textwidth}{}
 }
\caption{Same as Figure \ref{fig:periodogram}, but as a function of temperature instead of color, zooming on the stars with convective envelopes and splitting sources in different age bins into different panels. \teff\ is obtained from TESS Input Catalog. The 2d histogram shows the distribution of periods in this work, color coded by the number of sources in each bin, blue dots are from literature: Upper Sco \citep{rebull2018}, Pleiades \citep{rebull2016}, NGC 3532 \citep{fritzewski2021a}, Praesepe \citep{douglas2019} and NGC 6811 \citep{curtis2019a}. Black line is drawn at the period of 7 days, corresponding to half an orbital period from TESS. Note that the overdensity at that period that is apparent at older ages is artificial, most likely dominated by real rotators with longer period aliased to the harmonic period close to the half an orbit of TESS. Additionally, the sample in TESS may be more incomplete than the samples from the literature due to the magnitude limits.
\label{fig:periodogramage}}
\end{figure*}

As noted above, Figure~\ref{fig:periodogram} shows a clear gradient of rotation periods as a function of age and color, especially at its upper envelope (i.e., slower rotators). Indeed, separating the stars into individual age bins in Figure~\ref{fig:periodogramage}, it becomes more apparent the majority of stars of a given age populate a coherent, inverted horseshoe-shaped ``slow sequence" \citep[or ``I-type" sequence; see, e.g.,][]{barnes2010} in the color--period diagram that evolves smoothly with age.

Figure~\ref{fig:periodogramage} also compares the sequences we observed with those reported by previous authors for some well-studied clusters of known ages \citep{rebull2018,rebull2016,fritzewski2021a,douglas2019,curtis2019a}. In general there is excellent agreement at those ages where the previous samples exist, except for the slowest rotators, where our incompleteness for periods longer than $\sim$12~d becomes most apparent (see Section~\ref{sec:periods}).
At the youngest age bins, the distribution of rotation periods appears to be fairly complete (Figure \ref{fig:periodogramage}). There is an artifact in the catalog showing an excess of sources with a period of $\approx$7 days; as previously mentioned this is due to the systematics due to the orbit from TESS. This excess becomes particularly pronounced at older age bins, $>$100 Myr, but seems to be largely absent in the younger populations. This is most likely a signature of a periodic star, but having a dominant period longer than what is recoverable, as at older ages our period search is truncated at 12 days (see Section~\ref{sec:periods}).

The TESS sample reported here for stars in the large set of stellar associations from the Theia catalog \citep[][see Section~\ref{sec:theia}]{kounkel2020} significantly enlarges the overall sample of stars with rotation periods and ages, and importantly it also smoothly fills in the entire age range from $\lesssim$10~Myr to $\gtrsim$1~Gyr. This smoothly evolving and well populated slow sequence over such a large age range forms the basis for empirical gyrochronology relations, as we discuss in Section~\ref{sec:gyro}. 

\subsection{Rapid rotators in the color--period diagram}\label{sec:binaries}

Outside of the dominant slow sequence, Figure~\ref{fig:periodogramage} also shows at most ages a less coherent population of much more rapidly rotating stars with periods $<$2 days. 
These have been identified in early works as members of a so-called ``C-type sequence" \citep[see, e.g.,][and references therein]{barnes2010}. 
In this section, we consider the nature of the rapid rotators and their evolution with stellar age.

\subsubsection{Rapid rotators as binaries}

\begin{figure*}[t]
 \centering
 \epsscale{1.1}
\plottwo{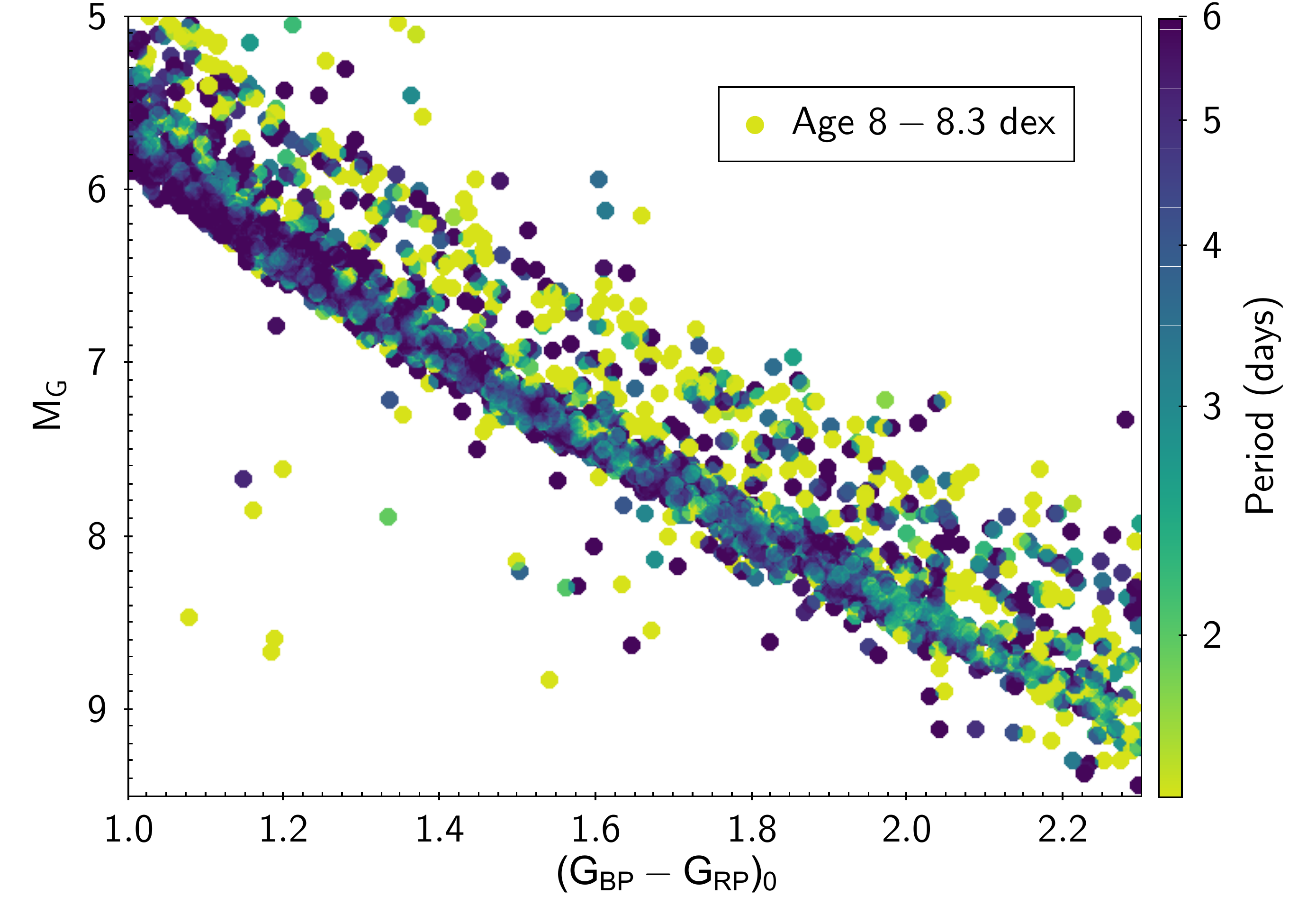}{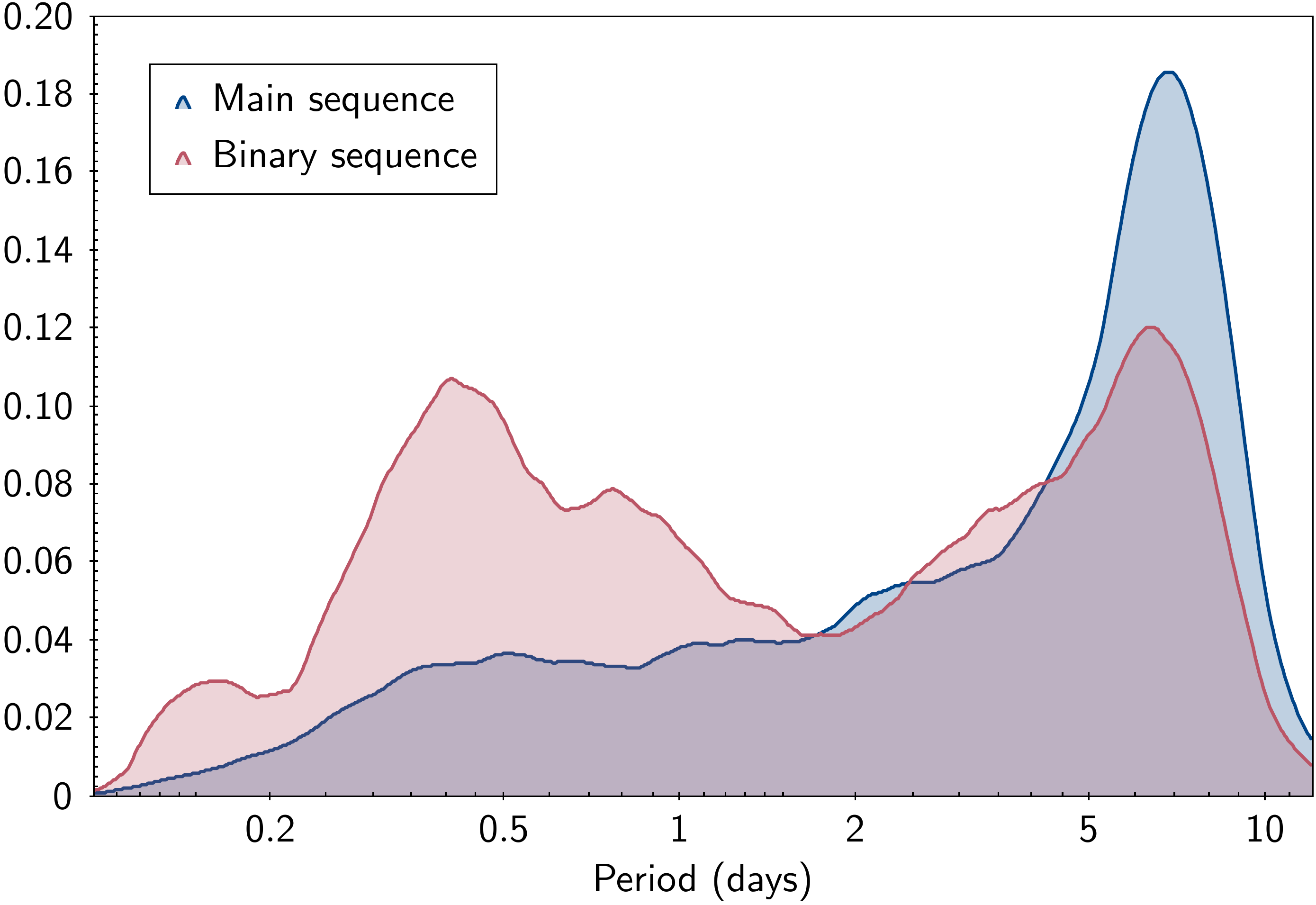}
\plottwo{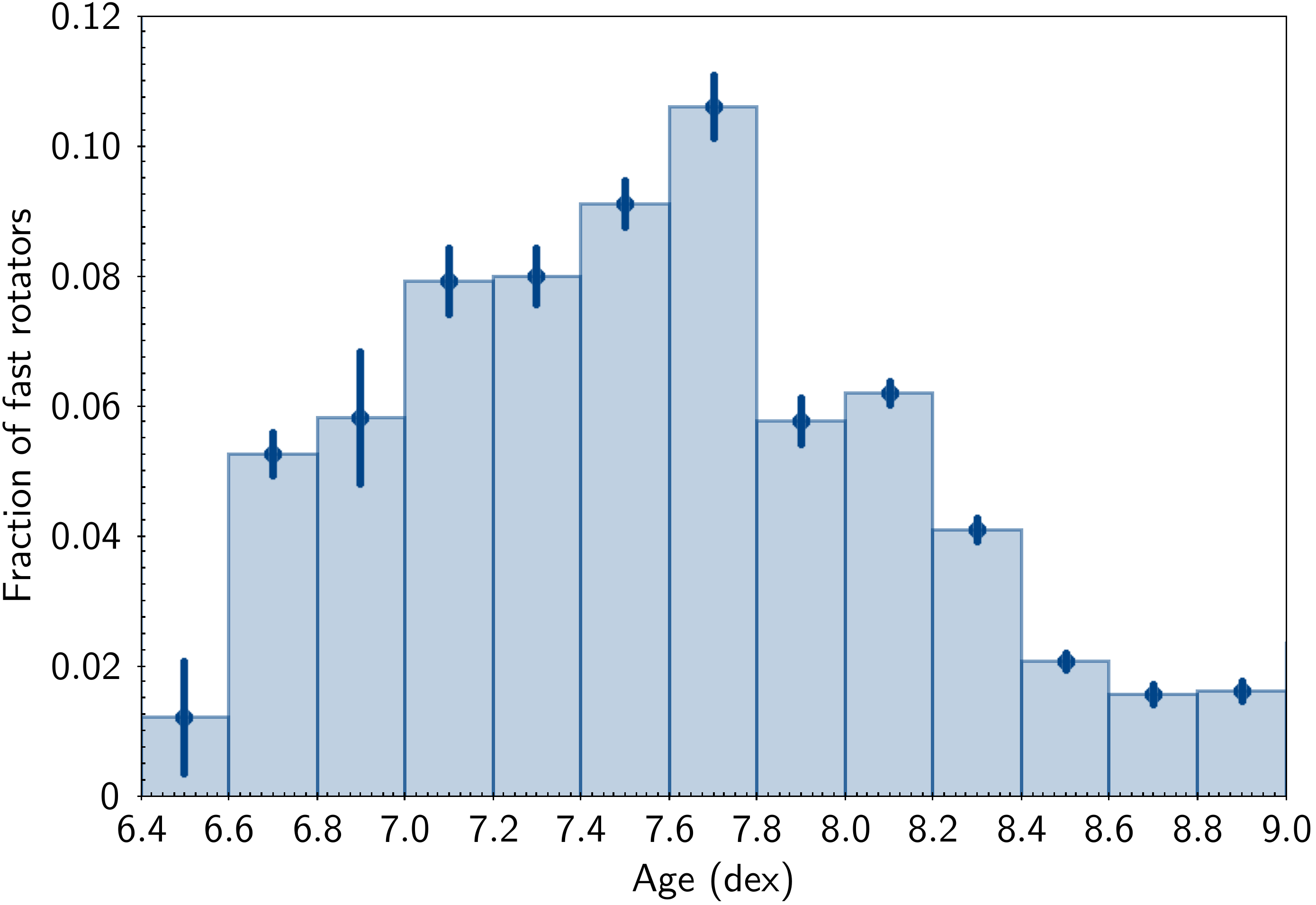}{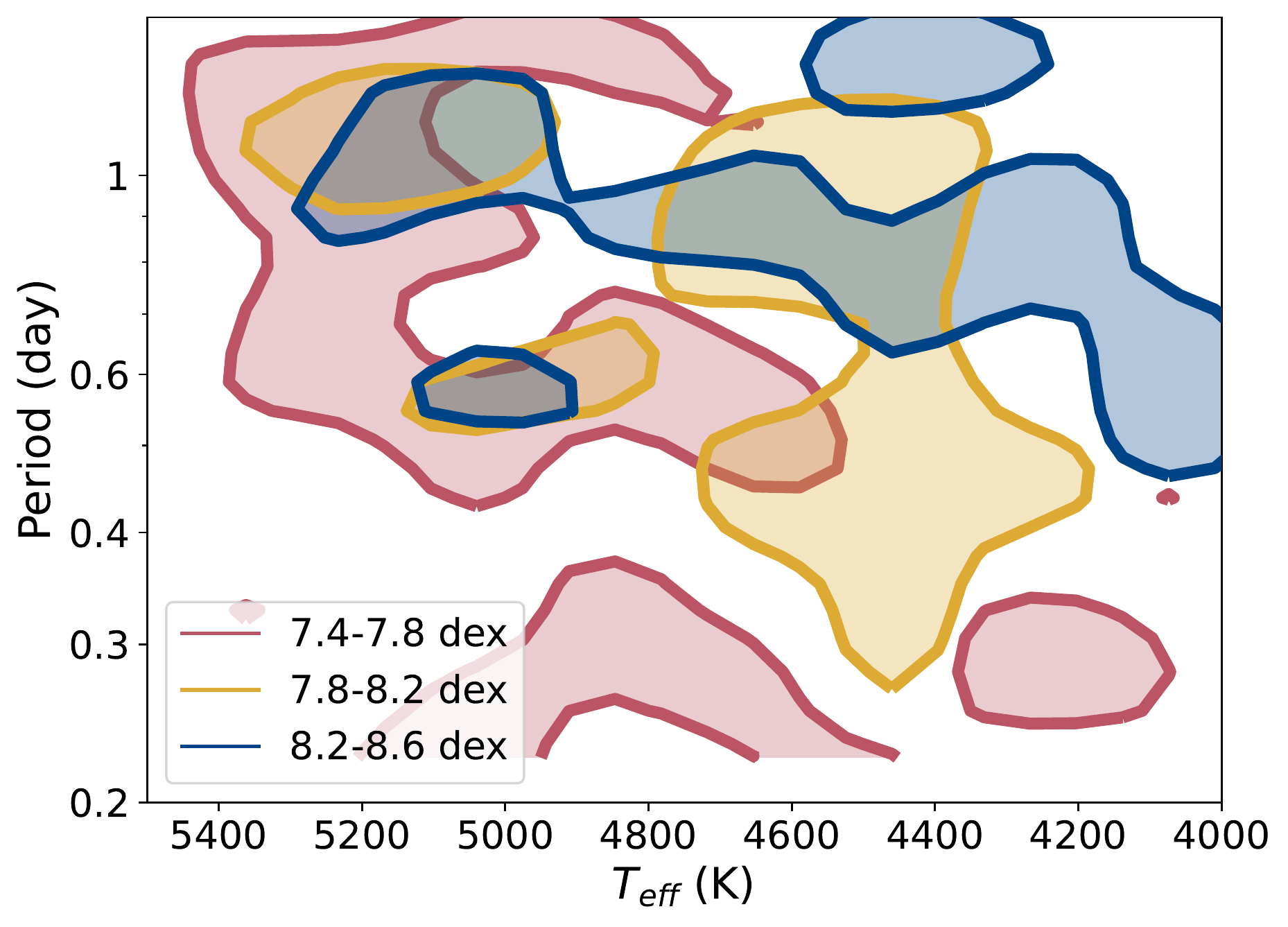}
 \caption{Top left: an HR diagram for the stars with ages between 8--8.3 dex; sources are color-coded by the recovered rotation period. Top right: a kernel density estimate showing the period distribution in the same age and color range for the sources located on the main sequence versus those found on the binary sequence. Fast rotators with periods shorter than 2 days are preferentially found on the binary sequence. Bottom left: Fraction of fast rotators (sources with extinction corrected colors 0.8$<(G_{\rm BP,0} - G_{\rm RP,0})<$2.2, with period $<$1 day), relative to all the stars (periodic or not) in the same color range. Bottom right: contour plot showing periods and temperatures of fast rotators as a function of age; as population ages, fast rotator sequence tend to shift towards longer periods and cooler temperatures.
 \label{fig:binaries}}
\end{figure*}

Recent studies with Kepler and K2 have uncovered a strong tendency for rapid rotators to be binary stars, presumably undergoing tidal interactions. For example, \citet{douglas2016}, \citet{Douglas:2017}, and \citet{simonian2019} all find that binaries dominate among FGK-type stars with periods faster than 2 days, and \citet{stauffer2018} find that binaries dominate among M-type stars with periods faster than 1.5 days. Indeed, comparing the stars in our sample that sit on the main sequence versus those on the binary sequence (Figure~\ref{fig:binaries}), we find that the binary sequence is heavily dominated by fast rotators with periods of 2 days and faster. 
Of course, not all true binaries sit on the binary sequence (i.e., binaries with small brightness ratios sit on the main sequence), 
and similarly not all of the unresolved binaries are tight enough to have experienced spinup to short periods. We also compare RUWE, but there is no significant difference in RUWE between fast and slow rotators.

\subsubsection{Evolution of the rapid rotators}

There is some evolution of the fastest rotators with age. For example, Figure~\ref{fig:binaries} examines their rough fraction relative to the total sample, selecting stars with colors of 0.8$<(G_{\rm BP} - G_{\rm RP})<$2.2 and period $<$1 day, relative to all the stars (periodic or not) in the same color range. We find that these very fast rotators are most common at ages of 40--60~Myr. There are approximately twice as many in this age bin compared to younger stars at $\sim$5~Myr, and very few are found at $\sim$3~Myr.
This is consistent to what has been previously observed in various young open clusters \citep[see, e.g.,][]{bouvier2014}, and has been interpreted as being due to the initial
releasing of ``disk-locked" stars \citep[e.g.,][]{affer2013} and their subsequent spin-up 
caused by core--envelope decoupling \citep[e.g.,][]{moraux2013}. 
There is also a decline of the fastest rotators at ages $>$60~Myr, which is at least partially driven by the overall decrease in recovered periodicity at older ages (Figure \ref{fig:fraction}) due to smaller starspot sizes (see Section~\ref{sec:spots}). 

It has been previously suggested that the evolution of fast rotators has some dependence on mass \citep{barnes2003}. Indeed, Figures \ref{fig:periodogramage} and \ref{fig:binaries} do show it. While the fast rotators tend to have a significant spread, there does appear to be a slight overdensity of stars that may resemble a fast ``sequence'' that becomes pronounced at the ages of 7.4 dex through 8.6 dex, and it persistently evolves towards cooler temperatures and longer periods for older stars. This fast sequence is significantly more diffuse than the slow sequence, and the relative density of sources along it is low, as such it is difficult to fully fit its evolution. But, eventually, it appears to merge into the slow sequence, leaving behind only a handful of faster rotators that remain outlying.

We return to a discussion of the rapid rotators in Section~\ref{sec:discussion}. For the purposes of exploring gyrochronology and angular momentum evolution of nominally single stars, in the following we restrict our analysis to ``slow sequence" stars with rotation periods longer than 2 days.

\subsection{Angular momentum evolution and gyrochronology}\label{sec:gyro}

As noted above, the stars found in the slow/I-type sequence show a strong dependence of rotation periods on age, with F, G, and early K type stars rotating increasingly slower as they age. In contrast, 
following the turn-over in the color--period diagram at a color of $(G_{\rm BP} - G_{\rm RP})\gtrsim 2.3$ or $T_{\rm eff} \lesssim 3800$~K (Figure~\ref{fig:periodogramage}), the M dwarfs appear to 
follow a very steep relationship such that the stars overlap each other in color--period diagram regardless of their age (Figure~\ref{fig:periodogram}), which makes it difficult to use periods alone to estimate their ages. Consequently, empirical gyrochronology relations have by necessity focused on relating rotation periods to age for mid-F to mid-K type stars. 

\begin{figure}[!t]
\includegraphics[width=\linewidth]{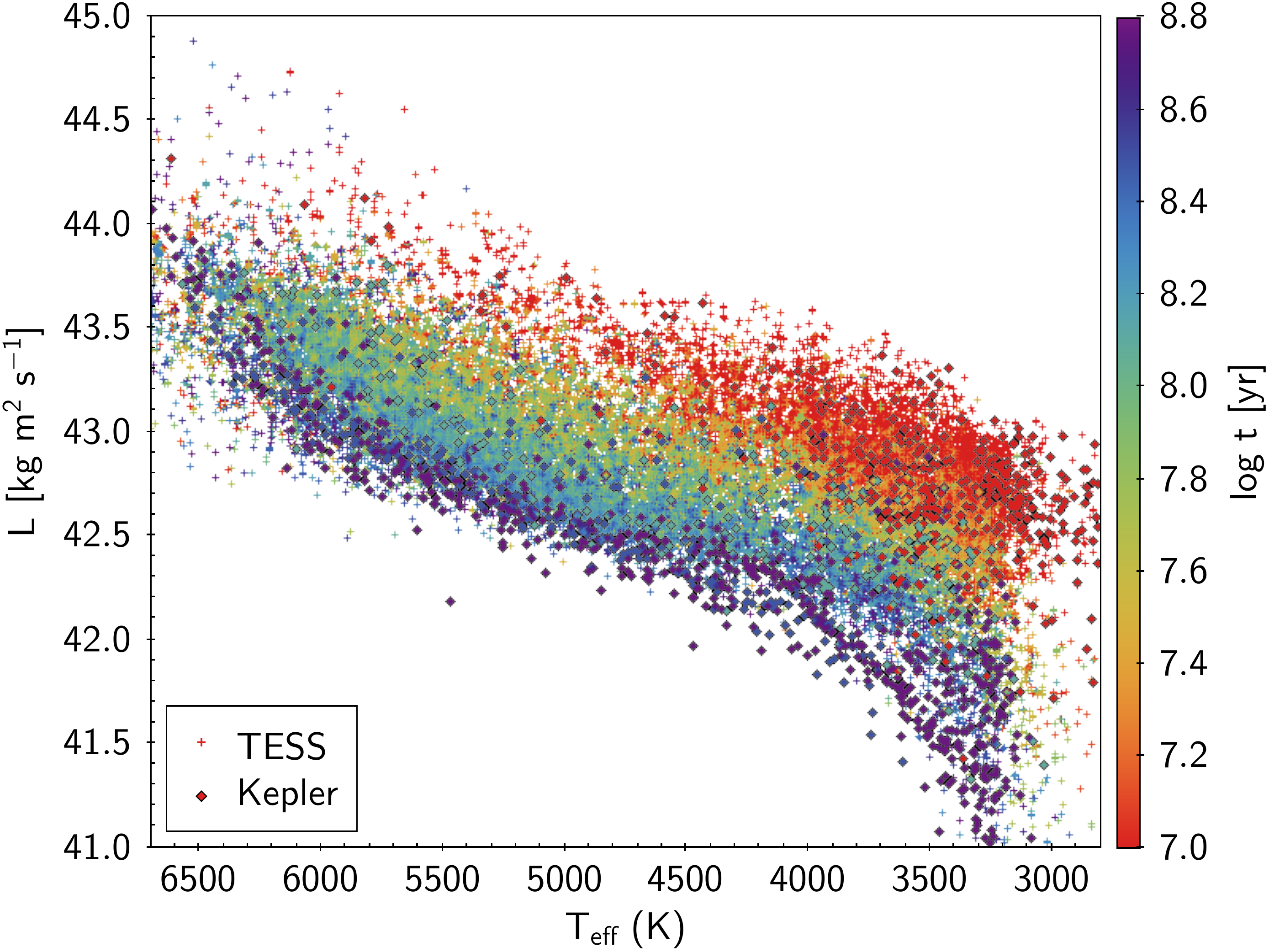}
\caption{Angular momentum of the full sample for ``slow sequence" stars with rotation periods longer than 2 days. Larger points are from the literature for clusters observed by Kepler. Note the significant change in $L$ as a function of age for the M dwarfs, despite the lack of a significant change in the rotation period (Figure \ref{fig:periodogram}).
\label{fig:angmom}}
\end{figure}

With the advent of precise Gaia parallaxes for the large catalog of stars with TESS rotation periods and Theia ages, we now have the opportunity to consider the stars' masses and radii along with their rotation periods. Thus, it is possible to estimate the total angular momentum ($L$) carried by each star (Figure \ref{fig:angmom}), using the Gaia-based radii and masses from the TESS input catalog \citep{stassun2019}. For simplicity, we calculate $L$ as for a solid-body rotator: $L=2/5MR^2\Omega$, for $\Omega = 2\pi/P$. We do not attempt in this exercise to account for possible core--envelope decoupling or other possible effects (e.g., metallicity).\footnote{Metallicity is implicitly included in the radius calculation, as it is derived from the Stefan--Boltzmann Law, thus requiring only apparent $G$, parallax, a color-based \teff, and a bolometric correction. Masses, however, were calculated solely from \teff, and so do not account for metallicity or stellar evolution \citep{stassun2019}.} Consequently, the quantitative $L$ values that we report here should be regarded as a proxy for the true angular momentum content that we will utilize for the purpose of relating the observables to age for gyrochronology applications.

We find that after excluding the fast rotators/binaries (i.e., excluding stars with periods shorter than 2 days; see Section~\ref{sec:binaries}), the age gradient is much more pronounced in $L$ space than it is in period space, indeed becoming flatter and monotonic with $T_{\rm eff}$ instead of inverse horseshoe-shaped (i.e, having a peak at $\sim$4000 K, and decreasing towards hotter and cooler stars). In addition, whereas rotation periods of slow-sequence stars do not evolve much during the pre-main-sequence (see Figure~\ref{fig:periodogramage}), the stars are in fact rapidly shrinking in radius, such that their $L$ content is rapidly decreasing and manifesting as a greater spread in $L$ than in rotation period. 
Although the turnover still occurs, it is found at somewhat cooler temperatures than in the period space, and the cool stars beyond the turnover are largely excluded by the cut of 2 days.

Using the angular momentum, $L$, we fit an empirical gyrochronology relation. We first find the average $L$ for a star of a given temperature and age. Each age bin is spaced by 0.1~dex, including sources within $\pm$0.1~dex of the bin center. We further bin \teff\ within each age bin, with \teff\ bins spaced by 50~K, and including stars within $\pm$200~K. We clean the sample in each 2~d period bin by finding a standard deviation in $L$ and excluding sources that deviate more than 1$\sigma$ from the median. We then recompute the median $L$ of the bin and iterate to convergence.
This process is applied using periods presented in this work up to ages of 8.5~dex. At older ages, our sample is highly incomplete due to missing longer periods (see Section~\ref{sec:periods}). As such, we tether our sample to the angular momentum of 670-Myr-old Praesepe \citep{douglas2019,cantat-gaudin2020} and 1-Gyr-old NGC~6811 \citep{curtis2019a}, thus extending the relations up to 1~Gyr.

We fit a simple polynomial relationship to the resulting data to estimate $L$ as a function of \teff\ and age as:
\begin{equation}\label{eqn1}
\begin{split}
\log L=a_0+a_1\logt+a_2(\logt)^2+a_3(\logt)^3+\\
b_0\log t +b_1\log t\logt+b_2\log t(\logt)^2
\end{split}
\end{equation}
\noindent where $\logt$ is the $\log_{10}$ of \teff\ in K, and $\log t$ is $\log_{10}$ of age in years.
This formalism is chosen to be reversible, such that, given a measurement of $L$, the age can be estimated via: 
\begin{equation}
\begin{split}
\log t=(\log L-a_0-a_1\logt-a_2(\logt)^2-\\a_3(\logt)^3)/(b_0+b_1\logt+b_2(\logt)^2)
\end{split}
\end{equation}

\begin{figure*}
\epsscale{1.1}
\plotone{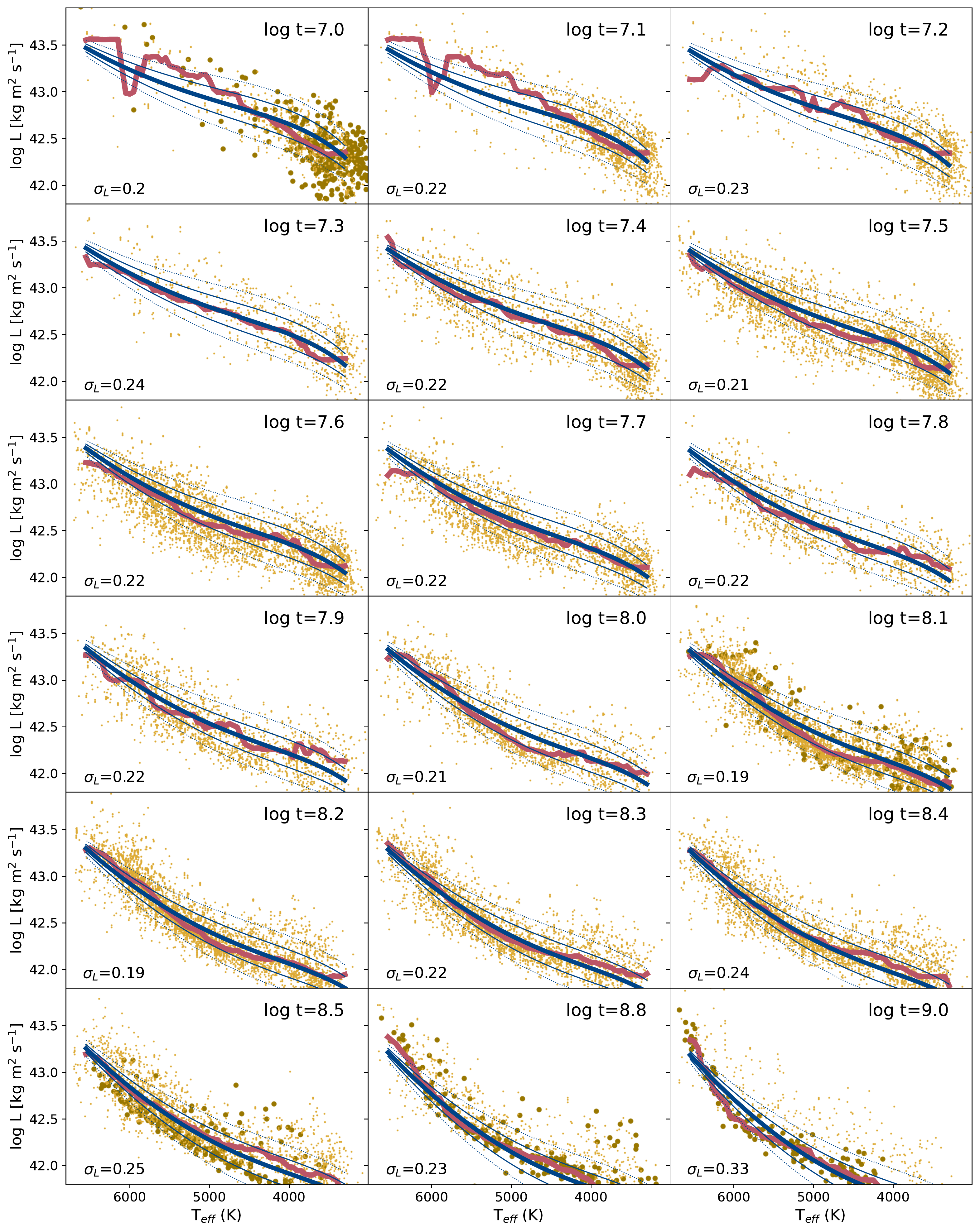}
\caption{A fit of the angular momentum $L$. Yellow dots show the data within each age bin; small dots are for sources with period measured in this work, larger dots are from literature: Upper Sco \citep{rebull2018}, Pleiades \citep{rebull2016}, NGC 3532 \citep{fritzewski2021a}, Praesepe \citep{douglas2019} and NGC 6811 \citep{curtis2019a}. The red line shows the median $L$ estimated within each age/\teff\ bin. The thick blue line shows the fit of $L$ for that age. Thinner blue lines are offset by $\pm$0.3 dex in age, and dotted lines are offset by $\pm$0.6 dex in age. $\sigma_L$ shows the typical scatter in $L$ relative to the fit in a given age bin. The range in $L$ corresponds to the valid domain for the fit.
\label{fig:fit}}
\end{figure*}

The fitted coefficients are presented in Table~\ref{tab:coeff} and the fit itself shown in Figure \ref{fig:fit}. The coefficients are strongly correlated, as such, the uncertainties for them are not provided, however, the number of coefficients cannot be reduced to minimize the correlation, as this leads to a poorer fit.

Separately, we also repeat a similar exercise with specific angular momentum $H \equiv L/M$. Note that the actual computation of $H$ does not invoke the stellar mass at all; we simply calculate $H$ using the stellar radius and rotation period alone. Interestingly, the scatter about the fitted relations is generally tighter in $L$ than in $H$. 

\begin{deluxetable}{ccc}
\tablecaption{Fitted coefficients for gyrochronology relations
\label{tab:coeff}}
\tabletypesize{\footnotesize}
\tablewidth{\linewidth}
\tablehead{
 \colhead{Coefficient} &
 \colhead{Value ($L$)} &
 \colhead{Value ($H$)}
 }
\startdata
$a_0$ & $-$3506.5298 & $-$3618.6016 \\
$a_1$ & 2697.8409 & 2712.4212 \\
$a_2$ & $-$677.8322 & $-$667.4448 \\
$a_3$ & 56.2650 & 54.0097 \\
$b_0$ & 113.3048 & 135.0571 \\
$b_1$ & $-$62.9542 & $-$74.6817 \\
$b_2$ & 8.7066 & 10.2867 \\
\enddata
\end{deluxetable}

The resulting relations are valid for nominally single pre-main-sequence and main-sequence stars (but not evolved stars) that follow the slow sequence, with 3000$<$\teff$<$6700~K, with rotation periods between 2 and 12 days, and with $\log L$(kg m$^2$ s$^{-1})>$41.8. This threshold restricts stars with ages of older than 300 Myr to $\gtrsim$4000 K.

\begin{figure}[!ht]
 \centering
 \includegraphics[width=\linewidth]{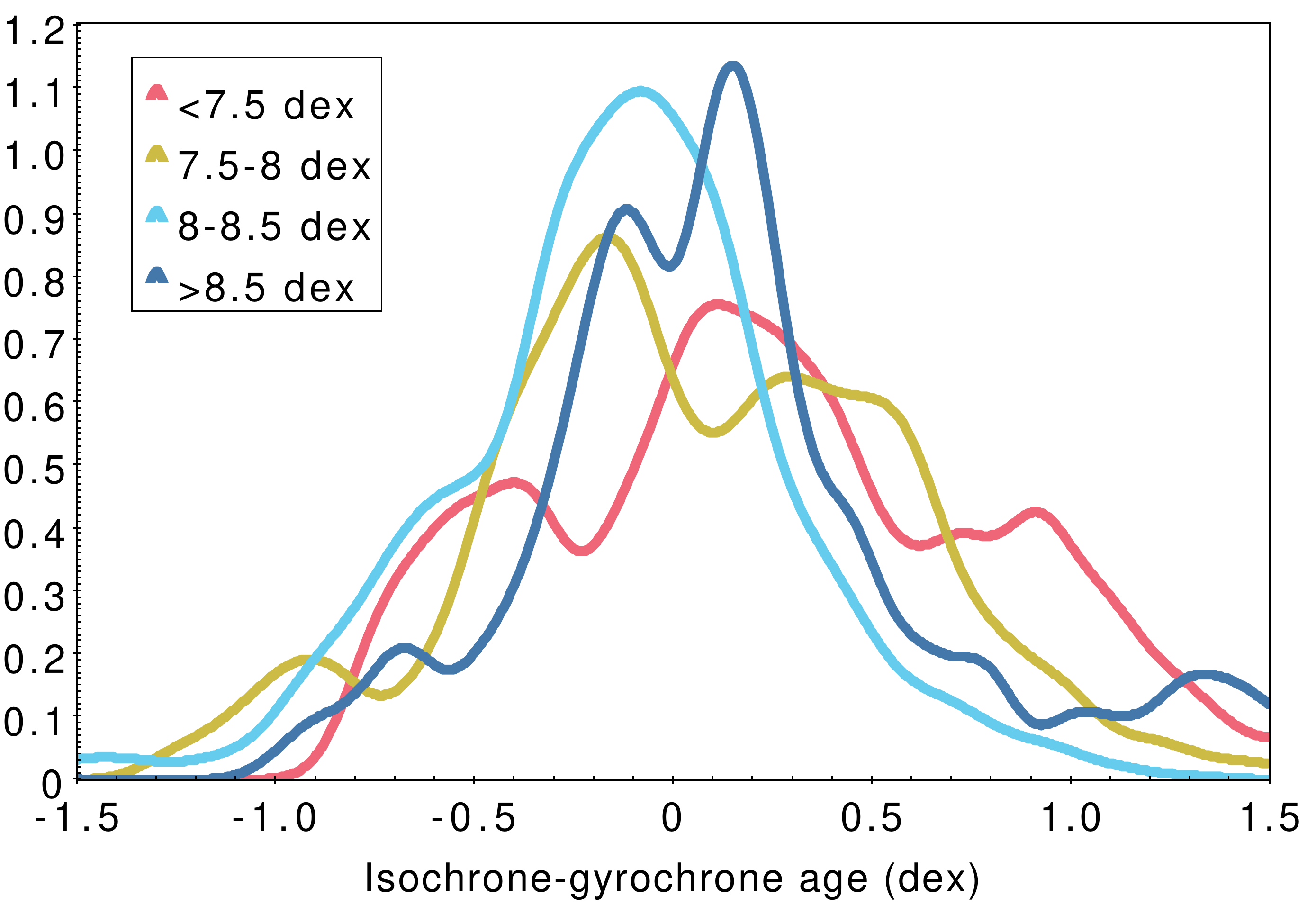}
 \includegraphics[width=\linewidth]{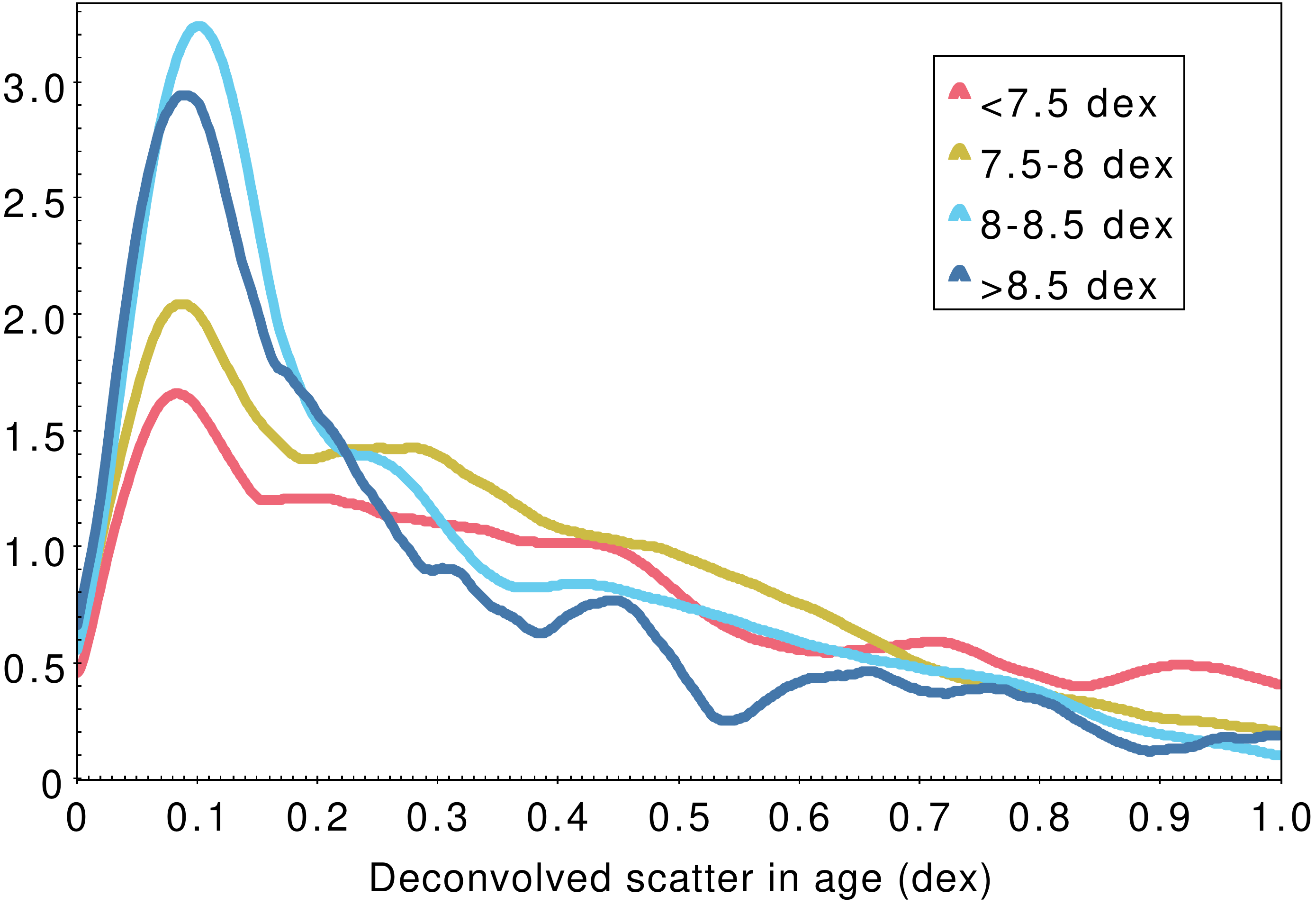}
 \caption{Top: A comparison of residuals between the age derived from angular momentum vs original ages derived from isochrone fitting, Bottom: same as above, but with the uncertainty in the isochrone age subtracted in quadrature from the difference. Different lines show the typical scatter at different ages: it is most pronounced at the younger age bins. Sources older than 8.5 dex are from Kepler, younger sources are from TESS.}
 \label{fig:scatter}
\end{figure}

To estimate the precision with which our fit enables measurement of gyrochronological ages, we compared its gyrochronological ages against the isochrone ages from the Theia catalog. 
The result is shown in Figure~\ref{fig:scatter}, which shows the distribution of age residuals (in the sense of gyrochronology-inferred age minus Theia nominal age, in dex) for the sample stars in broad age bins from $<$7.5~dex to $>$8.5~dex. The distributions have been ``deconvolved" by subtracting in quadrature the systematic errors intrinsic to the Theia-based age scale \citep[see][]{kounkel2020}. Therefore, Figure~\ref{fig:scatter} represents the relative age precision, not the absolute age accuracy, obtainable from our angular-momentum based gyrochronology relations. 
For the youngest age bin ($<$7.5~dex), the rotation-derived ages are largely uninformative beyond confirming their overall youth.
For the older age bins between 100 Myr and 1 Gyr, the 
distribution of the differences between the gyrochrone and the isochrone ages peaks at 0.1--0.2~dex, with a tail of significantly larger uncertainty. This is most likely due stars with ages $>$100 Myr begin to fully converge on a slow gyrochrone sequence in comparison to their younger counterparts.

\subsection{Mapping the evolution of starspot properties}\label{sec:spots}

Sources redder than the Kraft break have variability dominated by the rotation of the spots. The amplitude of variability decreases the closer the star in color is to the gap, as the size of the convective envelope shrinks, from typical variability of 0.5--1\% among young, low mass stars, to $<0.1$\% for the stars near the Kraft break. 
There is also a significant dependence of variability on the age of the star. Among the sources younger than 10 Myr, as many as 60--80\% of stars have a strong dominant rotation period (Figure \ref{fig:fraction}). However, fewer than $<10$\% of stars with ages older than 1 Gyr appear to be periodic. The cause of this is twofold. First: many older stars may have a rotation period longer than 12 days, which, so far, cannot be recovered with TESS.\footnote{Kepler light curves towards 1 Gyr old cluster NGC 6811 recovered periodicity of 171 stars out of 203, resulting in a 84\% fraction} But also, the overall amplitude of variability also decreases with age, as the activity is decreased, potentially making periodicity harder to recover \citep{Morris2020}.

A simplistic formula for translating the observed amplitude of variability into a spot coverage fraction $f_{\rm spot}$ uses the Stefan--Boltzmann relation, and can be written
\begin{equation}
f_{\rm spot}=2\sigma_{\rm var}/(1-T_{\rm spot}^4/T_{*}^4),
\end{equation}
where $T_*$ is the temperature of the unspotted photosphere, $T_{\rm spot}$ is the temperature of the spot, and $\sigma_{\rm var}$ is the standard deviation of the normalized light curve.

This formula assumes that $2\sigma_{\rm var}$ corresponds to the total amplitude difference in flux between the time a single spot is facing toward us versus the time when it is facing away from us. Of course, there is no reason to expect a rotating star to have a single spot (group) on its surface, and the general problem of inferring a starspot map given only a light curve is inherently under-constrained \citep{Basri_2020,Luger_2021}. In reality, the observed amplitude of the spot-induced signal depends on whether any asymmetry is present in the longitudinal surface brightness distribution of the star. A star heavily covered with starspots shows little light curve variation if the observed ratio of dark to bright regions remains constant over each rotational cycle. Our approximation for the spot coverage fraction $f_{\rm spot}$ should therefore be interpreted as a suggestive number, but not one with a solid quantitative foundation.

To use Equation 3 to infer spot filling fractions from the photometric amplitudes of the periodic variability in each light curve, we must adopt an assumed spot-photosphere temperature contrast (i.e., a value for $T_{\rm spot}/T_{*}$). We follow \citet{berdyugina2005} and \citet{herbst2021} in making $T_{\rm spot}$ and explicit function of T$_{*}$: $T_{\rm spot}=(-3.58\times10^{-5}T_*^2+1.0188T_*-239.3)$ K. While we lack directly measured spectroscopic \teff\ measurements for the bulk of the sample, we adopt \teff\ from TESS Input Catalog \citep{stassun2019} as $T_*$, and then use the equation above to predict the associated $T_{\rm spot}$. With this approach, we find $T_{\rm spot}$ = 4530 K and 2440 K for stars with $T_{*}=6000 K$ and $3000K$, respectively. 

\begin{figure}[!t]
\includegraphics[width=\linewidth]{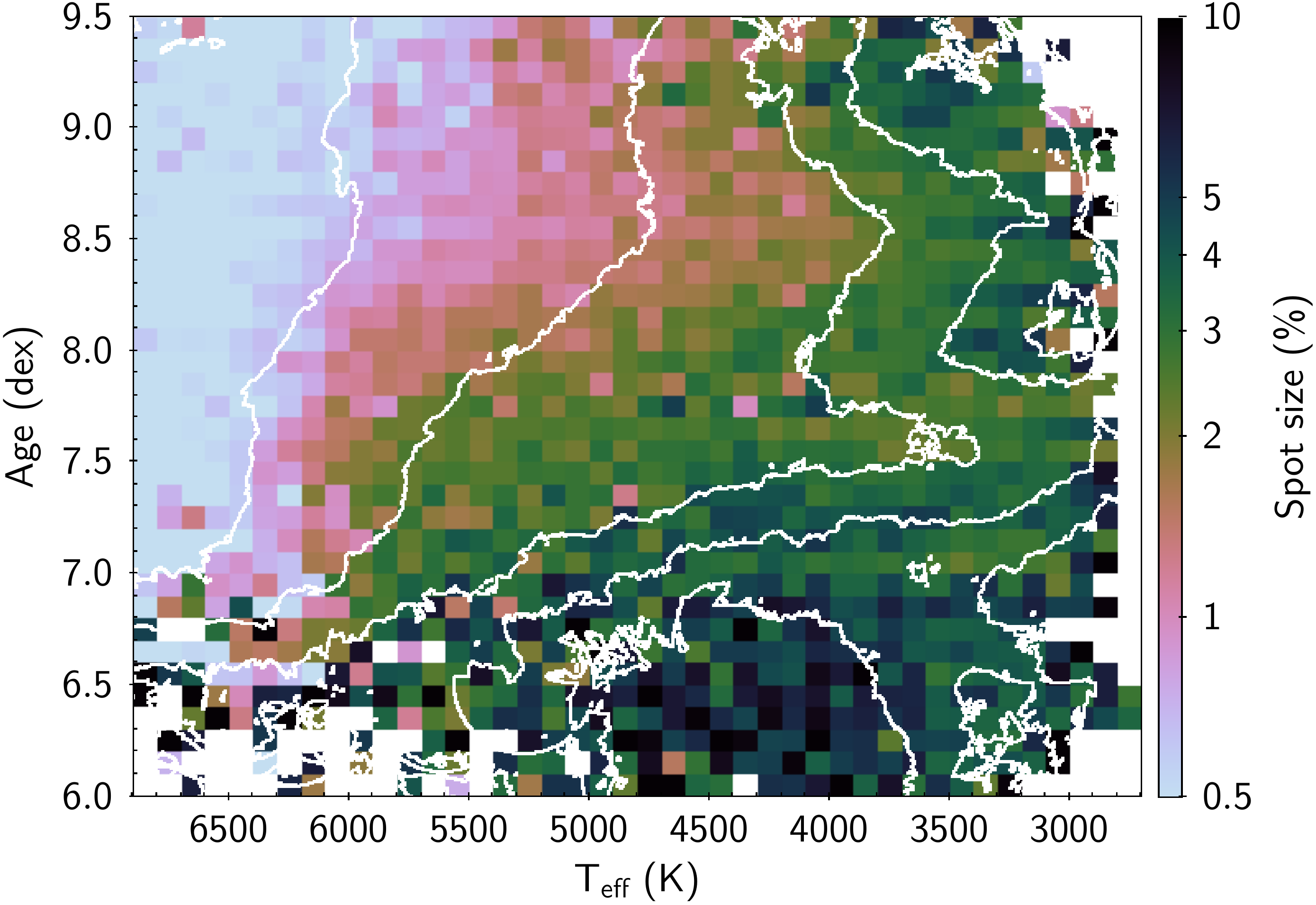}
\includegraphics[width=\linewidth]{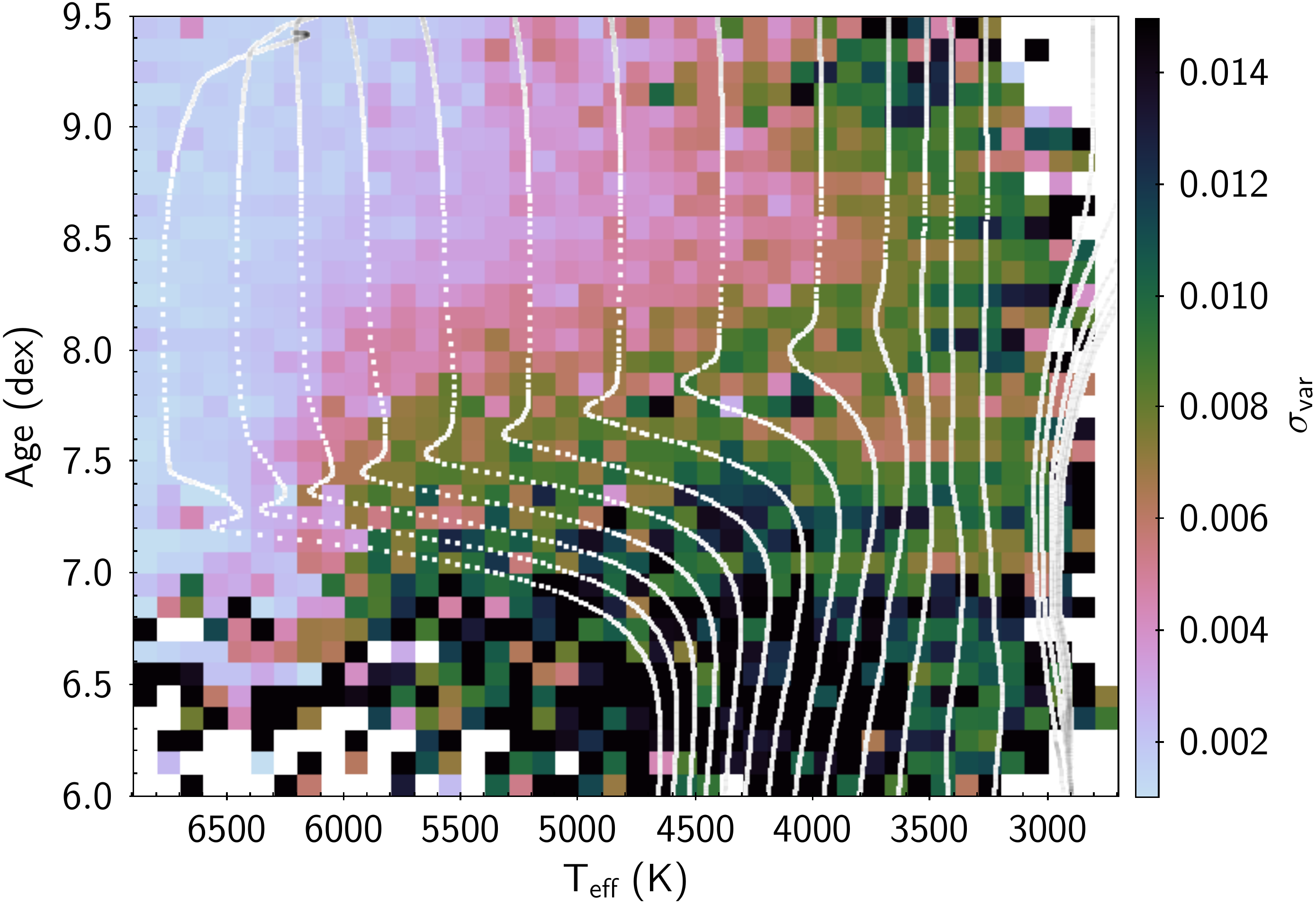}
\caption{Top: Median spot fraction size at different temperature and age bins. The white contours correspond to the levels of 1.0, 2.6, 4.5, 5.9, 7.3\%. Bottom: Same plot, but the color corresponding to the standard deviation of the light curve ($\sigma_{\rm var}$), with \citet{baraffe2015} isochrones plotted on top of it in white.
\label{fig:spotsize}}
\end{figure}

The resulting median spot fraction for the sample is shown in Figure \ref{fig:spotsize}. Indeed, we see a significant gradient in $f_{\rm spot}$ as both a function of age and temperature. The decrease in spot sizes is pronounced most strongly at two periods: after a pre-main-sequence star begins to develop a radiative core, and later, shortly after it settles onto the main sequence. For a solar-type star, we see a typical spot fraction of $\sim$5\% at the beginning of its life, decreasing by more than an order of magnitude beyond 1 Gyr. Such spot sizes are consistent with those observed by Kepler towards various young clusters \citep{Morris2020}.

A mid-M dwarf initially has $f_{\rm spot}\sim$10\%, modestly decreasing to $\sim$7\% after 1 Gyr. As such stars remain fully convective throughout their life, they are able to sustain large spots even at a relatively advanced age (which is only a small fraction of their total lifetime).

A young 4000 K star appears to have a typical spot size of $\sim$7\%. This is similar to the spectropolarimetric spot filling factor measurement of 4.5--6\% of the young T Tauri star DN~Tau \citep{donati2013}. However, this is significantly lower than 80\% observed for LkCa 4 \citep{gully-santiago2017}, obtained through fitting a two-temperature model to the H and K band spectra. Similarly, it is inconsistent with the spot coverage of 50\% of cool dwarfs in Pleiades through fitting TiO bands \citep{fang2016}. The difference may be systematic---it is possible that these spectroscopic studies preferentially track smaller spots that are more uniformly distributed along the photosphere. Such spots would not cause significant variability as both sides of a star would be equally spotted. On the other hand, \citet{donati2013} reconstruct only a singular large spot, which offers a better match to our underlying assumptions.

But, while a large spot filling fraction is far from typical in our sample, we do observe some individual stars that do have inferred $f_{\rm spot}$ as high as $>30$\%. Among young 4000 K stars, only $\sim$3\% of them appear to have such large spots. Some of these sources with large $f_{\rm spot}$ may be attributable to the remaining eclipsing binary stars that cannot be easily filtered out from the data. However, known spectroscopic binaries \citep{price-whelan2020}, or sources with large RUWE, or the sources on the binary sequence do not appear to dominate this large $f_{\rm spot}$ tail.

\section{Discussion} \label{sec:discussion}

\subsection{On the ability to use gyrochronology in the field}

There have been many attempts to construct theoretical models and empirical relations describing the evolution of stellar rotation \citep[e.g.,][]{barnes2003, barnes2007, barnes2010, angus2019, spada2020}
and angular momentum \citep[e.g.,][]{vanSaders2013, Matt2015, Garraffo2018}. 
The most popular parameterization for empirical gyrochronology, following \citet{barnes2007}, assumes that rotation's dependencies on mass (via $B-V$ color) and age are separable, such that stars already converged on the slowly rotating sequence spin down continuously with a common braking index. \citet{mamajek2008} report a typical age precision of 0.05 dex based on a fit to the rotation periods of FGK-type members of young canonical open clusters (e.g., $\alpha$~Per, Pleiades, M34, Hyades, spanning 80--700 Myr) to describe the mass/color dependence, with the Sun anchoring the age dependence (i.e., the braking index).

One of the limitations of this approach is that young clusters show stars with the same age and mass spinning at a range of periods. For example, single K dwarfs in the Pleiades, excluding those rotating more rapidly than $<$2 days, span a factor of five in period (and therefore angular momentum). This degeneracy places a limit on the age precision attainable with gyrochronology at particular masses and ages that are not yet well defined. This is presumably a relic of the initial angular momentum distribution, which some approaches account for \citep[e.g.,][]{barnes2010, Matt2015}; however, our relation instead tracks the broad trend defined by our large sample of stars and structures. 

In addition to this fundamental limitation due to the convergence process and timescale, all models proposed to date have also failed to describe the detailed spin-down behavior for FGK stars on the slowly rotating sequence. Magnetic braking appears to effectively shut down in the later main-sequence as stars approach Rossby numbers of $Ro \approx 2$ \citep[e.g., occurring at $\sim$2 Gyr and $\sim$4 Gyr for F and G dwarfs, respectively]{vanSaders2016, vanSaders2019, Hall2021}; this causes similar stars to pile up in rotation at a range of old ages \citep{David2022}. Earlier on in their main sequence evolution, K~dwarfs appear to temporarily stall for an extended period of time at ages of 0.6--1.5~Gyr, which also piles up stars over this age range before spin-down resumes \citep[e.g.,][]{Agueros2018, curtis2019, curtis2020}. 

These deviations from a pure Skumanich-style $P_{\rm rot} \propto t^n$ relation \citep[e.g.,][]{skumanich1972, barnes2007, angus2019} greatly reduce the true age precision attainable with rotation at certain masses and ages. It also has impacted the accuracy of most calibrations. For example, \citet{curtis2020} showed that this class of model predicts ages for the three components of 36~Oph that vary by a factor of two, while the gyrochronal age of 61~Cyg A and B are half that of their interferometry-constrained isochronal ages. In other words, although \citet{mamajek2008}, \citet{barnes2007}, and others have achieved seemingly high gyrochronal age precisions with their models based on the calibration data available at the time, newer data have proven the underlying assumptions of these models to be invalid, and the resulting ages to be quite inaccurate for low-mass stars in particular.

While the sample studied here is younger than the age when K dwarf stars appear to temporarily stall ($<$1~Gyr), \citet{curtis2020} suggested that G stars might also spin down very little between 100 and 200 Myr (8.1 and 8.3~dex in log age). This is on par with the typical 0.1--0.2~dex age uncertainties found for our empirical fit. Comparing our $L$--age relation with data for the Pleiades \citep{rebull2016a}, we find a rather poor fit across 4000--6000~K (see the $\log t = 8.1$~dex panel of Figure~\ref{fig:fit}): although the average rotation age for the ensemble is approximately correct, the warmer stars appear too young ($\sim$7.5~dex or 30~Myr) and the cooler stars appear too old ($\sim$8.6~dex or 400~Myr). It is possible that our empirical fit is simply not flexible enough to describe the evolution of $L$ across time and \teff. 

By 300--400~Myr, our relation appears to accurately describe the shape of the $L$--\teff\ distribution. However, we found that ages for the $\sim$350~Myr clusters NGC~3532 and NGC~7092 come out at 800--900~Myr ($\sim$2.5 times too high). This is, perhaps, partially due to inaccurate interstellar reddening corrections affecting the calculation of \teff, mass, and radius, but might also signify a systematic bias in the ages derived with our fit. We note that inaccuracies in interstellar reddening will most strongly affect the ages for the warmer late-F and early-G dwarfs in the steeper portion of the $L$--\teff--age space. For example, perturbing $A_V$ by $\pm$0.1 mag would cause 6000~K members of the $\sim$700~Myr Praesepe cluster to appear to be 100~Myr or over 1~Gyr if $A_V$ is under- or overestimated.

\begin{figure*}[!ht]
\includegraphics[width=\linewidth]{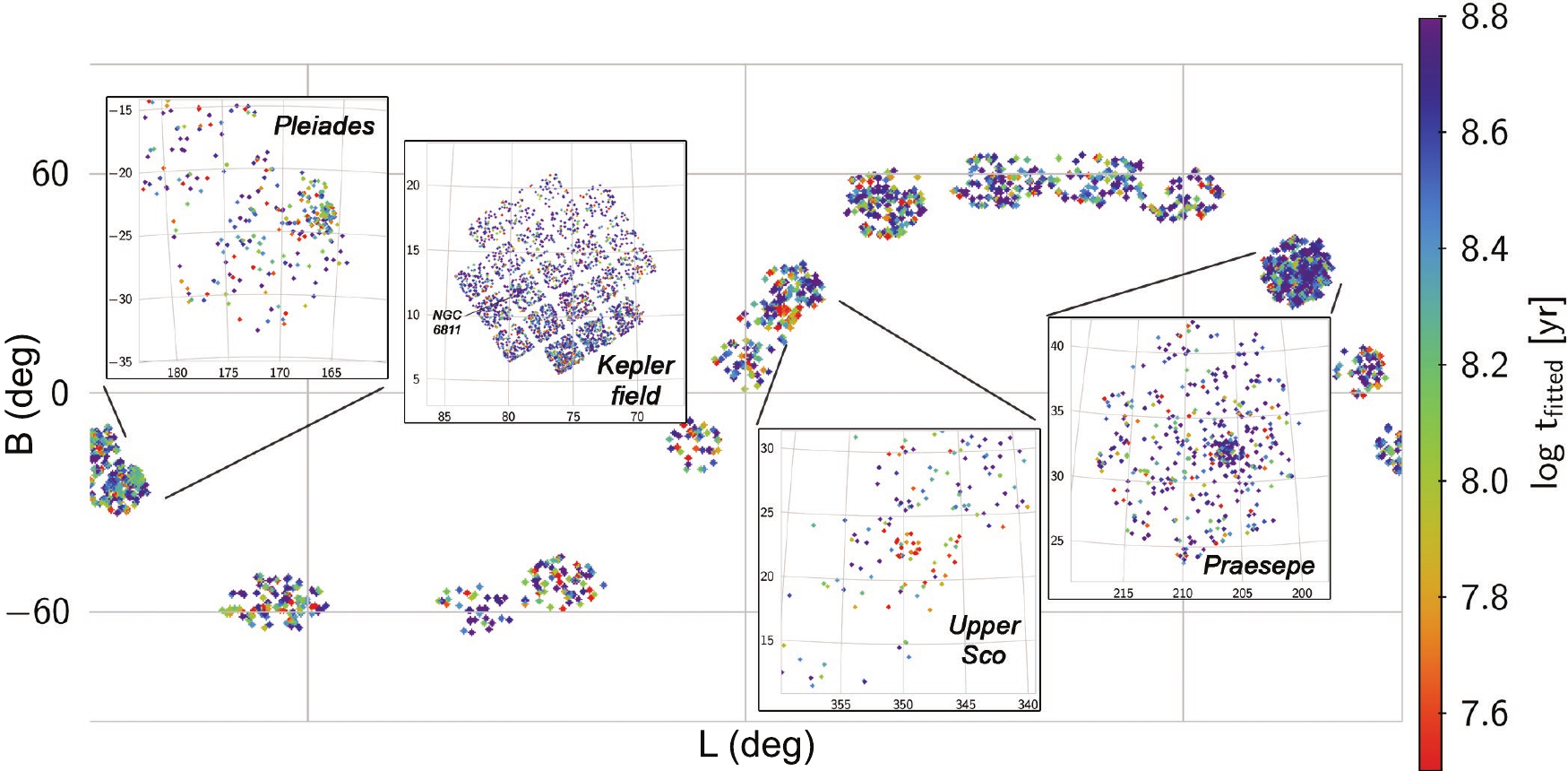}
\caption{Sky map of sources in \citet{mcquillan2014} and \citet{reinhold2020} catalogs in Galactic coordinates that satisfy the criteria of fit. All the stars are color-coded by their estimated age, as inferred from the empirical gyrochrone relations given in equation 2. The large panels zoom in on certain fields containing notable clusters and the Kepler field (not drawn to scale).
\label{fig:reinhold2020}}
\end{figure*}

Such issues ubiquitously plague all current approaches to age-date stars with rotation. 
Despite these challenges, Figure~\ref{fig:angmom} does show a continuous evolution of $L$ versus \teff\ over time that allows for ages to be estimated with a typical uncertainty of 0.1--0.2~dex.\footnote{It will be necessary in the future to cross-examine our empirical fit with other approaches in the literature \citep[e.g.,][]{spada2020} to determine if there is real improvement in the ages attained \citep[e.g., following][instead of taking each model's claimed uncertainties at face value]{Otani2022}, but this is beyond the scope of our present study.}
Thus it is now possible to estimate a prior on individual stellar ages inferred from angular momenta, which can now be measured across nearly the entire sky. At the moment, few rotation periods have been reported for field stars in TESS; however, rotation periods from Kepler \citep[e.g.,][]{mcquillan2014, Santos2021} and K2 \citep{reinhold2020, Gordon2021} are now available for large numbers of stars. We apply the aforementioned cuts to the data on period and fitted age, and we also require stars with radius $<$1.2 \rsun, to exclude subgiants, as well as exclude sources on the binary sequence. This significantly limits the catalogs to only 7\% of their original size, as most of the remaining stars have longer periods. We examine the resulting age distribution of the subset where the age fit can be considered valid in Figure~\ref{fig:reinhold2020}.

We find that the stars that can be associated with various young open clusters that have been included in these catalogs stand out from the field, and that their average age well matches the true age of the cluster. As such, with a large enough catalog of rotation periods, in future it may be possible to identify groups of stars that have formed together even after the point where a population is kinematically coherent as a moving group. Additionally, combined with completeness information as a function of age (e.g., Figure \ref{fig:fraction}), it may help with with better deriving the star forming history of the solar neighborhood.

Furthermore, combining priors on age derived via gyrochronology with the priors derived from kinematics \citep{angus2020, LucyLu2021} or from chemical clocks \citep[e.g.,][]{daSilva2012,Spina2018, moya2022} may allow the derivation of more precise ages even for individual stars.

\subsection{Implications for models of angular momentum evolution}

The ability to make empirical measurements of angular momenta for large numbers of stars with empirical age estimates from a few Myr to a Gyr offers a new opportunity to empirically inform theoretical models of angular momentum evolution. Indeed, long-standing questions persist regarding the respective roles of the different mechanisms by which stars of different masses/structures can shed angular momentum, and how the respective roles of those mechanisms may change as the stars evolve. 

\begin{figure}[!ht]
 \centering
 \includegraphics[width=\linewidth]{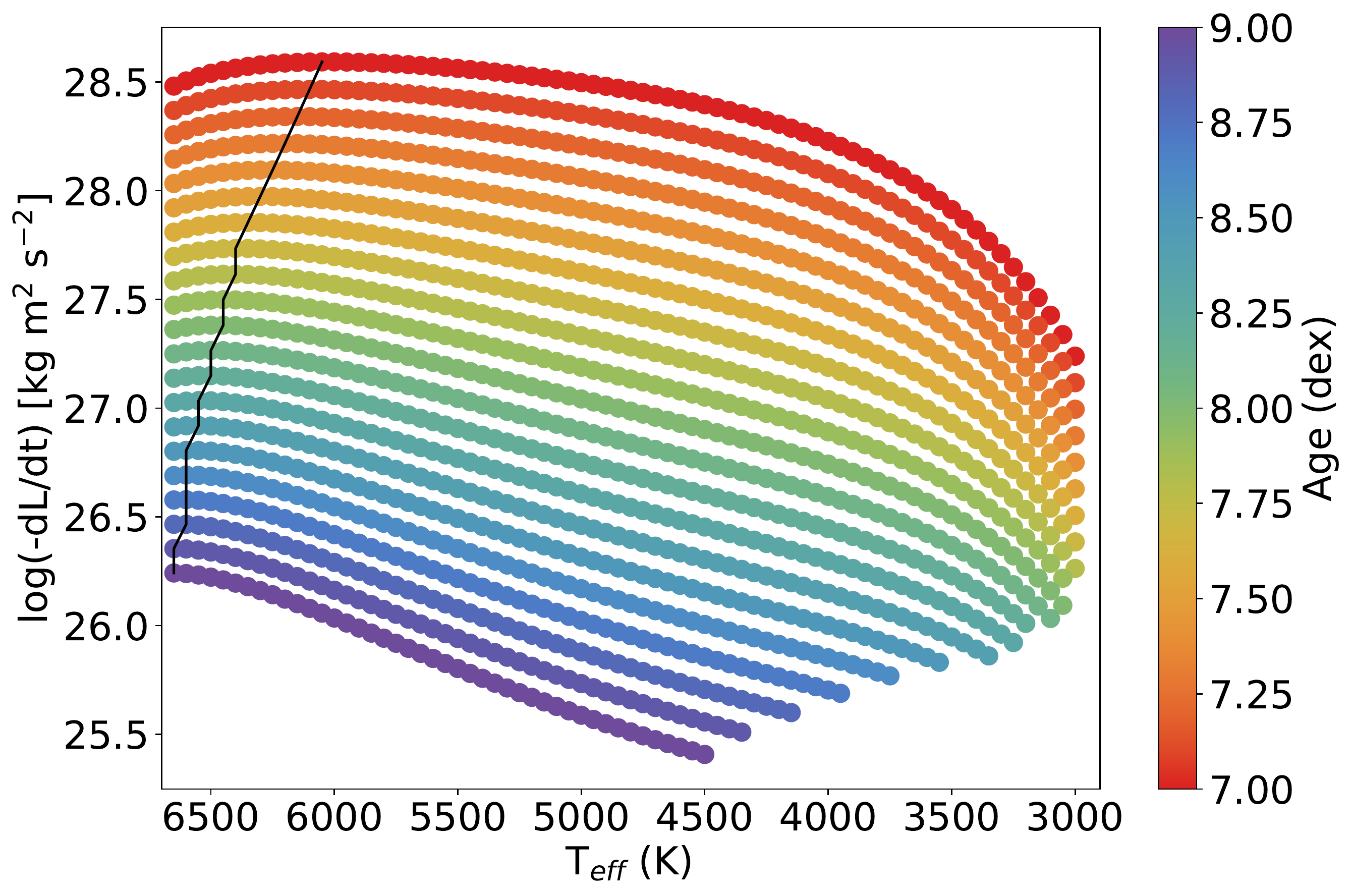}
 \includegraphics[width=\linewidth]{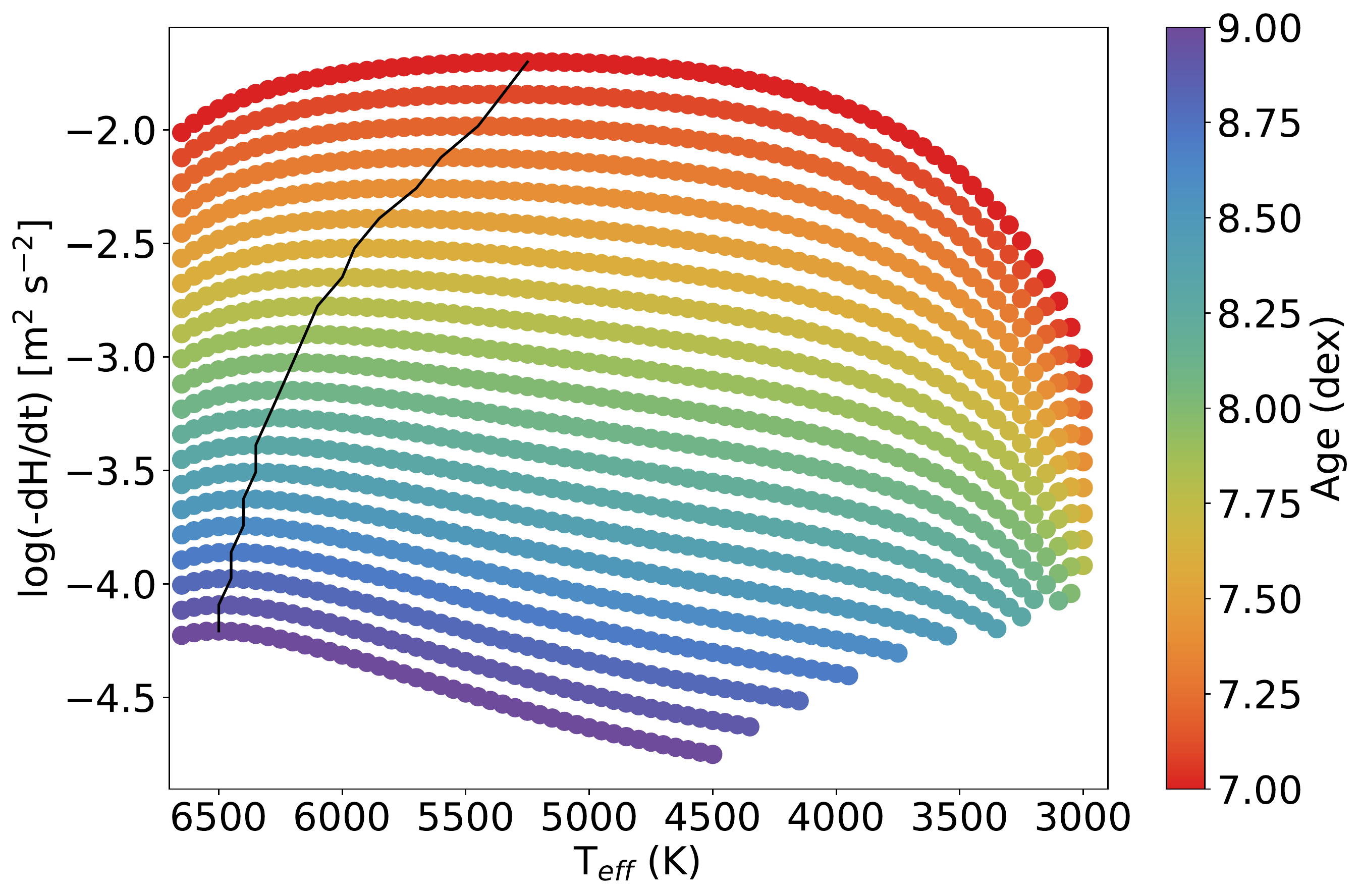}
 \caption{Rate of loss of angular momentum $L$ (top) and specific angular momentum $H$ (bottom), as a function of temperature and age, assuming the empirical relation in Figures~\ref{fig:fit}. The black line tracks the temperature at which the greatest loss-rate occurs for a given age.}
 \label{fig:dlh}
\end{figure}

As a first examination utilizing our large catalog, we examine the {\it rate of change} in both $L$ and $H$ as a function of age and \teff\ through taking a derivative of our empirical relations (Equation~\ref{eqn1}) with respect to time. The results are represented in Figure~\ref{fig:dlh}.

We make the following observations. First, angular momentum loss is a ubiquitous feature of mid-F to mid-M type stars at all ages (up to the 1~Gyr age limit of the sample investigated in this work). Second, the rate of angular momentum loss changes precipitously during the first Gyr, decreasing by up to 3 orders of magnitude during this time. Third, the angular momentum loss rate is greatest for stars of roughly solar mass and above at all ages, with the maximal angular momentum loss rate shifting to the slightly more massive stars as they age. Finally, M-type stars exhibit a significantly reduced rate of angular momentum loss at all ages compared to even slightly more massive stars, and this difference appears to be somewhat discontinuous across the fully convective boundary, suggesting that the dominant angular momentum loss mechanism(s) may be qualitatively different for fully convective stars and not merely quantitatively weaker.

We also make consider rapid rotators. We have seen in Section \ref{sec:binaries} that they tend to be binaries. Although an initial assumption might be that they are tight, tidally locked systems, Figure \ref{fig:binaries} shows that rapid rotation occur ubiquitously across the binary sequence, including unresolved systems spanning separations of several au (the pixel scale for Gaia of 0.06'' corresponds to 30 au at the distance of 500 pc). We also find that there is an apparent evolution of the rapid rotators with age, such that they slow their rotation periods on timescales of $\sim$1 Gyr and eventually approach the gyrochronological ``slow sequence''. This suggests that unresolved binaries---across large scales of separations---emerge from the formation process as rapid rotators and then, because they are not all tidally locked, the majority of them are able to spin down. While we cannot answer precisely how the youngest binaries come to be rapidly rotating even if they are not close enough to tidally interact, the observational evidence presented here appears to require such a scenario.

Confronting models of angular momentum evolution with these empirical trends will be an important next step in understanding the physics of angular momentum loss, a capability that is now possible given the availability of empirical angular momentum measurements together with fine-grained, empirical age estimates for large numbers of stars in the field.

\section{Conclusions} \label{sec:conclusion}

We generate light curves from TESS FFI data for stars within 500~pc and with known ages from the Gaia-based Theia catalog \citepalias{kounkel2020}. Our resulting catalog of $\sim$100,000 periodic variable stars contains primary, and in some cases secondary, periods up to a maximum of 12 days. The zoo of variability classes represented in this catalog is illustrated in Figure~\ref{fig:periodogram}.

We find a clear separation in the period distribution between stars with convective and radiative envelopes occurring at \teff$\sim$6700~K, corresponding to the Kraft break. In stars with outer convective envelopes, we find a strong correlation of variability with age. Up to 80\% of the youngest stars with age $\lesssim$10~Myr appear to be periodic, whereas only $\sim$10\% of stars with age $\gtrsim$1~Gyr can be recovered as such. Partially this decrease is driven by these older stars having rotation periods longer than periods that can be measured within a single sector of TESS, and partially it is driven by the decrease of starspot sizes, resulting in variability amplitude lower than can be securely detected. In addition to age, spot size also strongly depends on \teff, with lower mass stars being more heavily spotted on average. However, the recovered spot filling fraction may be underestimated in younger stars, as we are not sensitive to small and uniformly distributed spots.

A significant fraction of stars are fast rotators. They are most common at ages of 40--60 Myr and less prominent at the oldest and younger ages. Initially this is due to the stars spinning up as they settle onto the main sequence. Eventually, magnetic braking allows them to spin down to have rotation periods that place them on the so-called ``slow/I-type sequence" that has been found in previous works to be useful for gyrochronology \citep[see, e.g.,][and references therein]{barnes2007,barnes2010,mamajek2008}. 

We observe a clear and quite well defined slow sequence at all ages considered in this study, from 10~Myr to 1~Gyr, confirming previous results \citep[e.g.,][]{rebull2018,douglas2019, curtis2019a}. We also corroborate previous findings that rapid rotators with periods faster than two days are dominated by likely binaries, presumably due to tidal interactions that spin up the stars and maintain the rapid rotation \citep[see, e.g.,][]{douglas2016,Douglas:2017,simonian2019}. 

Limiting the sample to only sources on the slow/I-type sequence, with periods between 2--12 days, we develop an empirical gyrochronology relation of angular momentum evolution of pre- and main-sequence stars with convective envelopes. This fit allows angular momentum to be used to estimate ages from as young as 10~Myr and up to 1 Gyr. Importantly, through the use of angular momentum as opposed to rotation period alone, these empirical relations enable estimating ages of not only FGK-type stars but also M-type stars, due to the angular momentum distribution being much smoother, simpler, continuous and monotonic as compared to the rotation period distribution. As a result, we are also able to begin tracing in fine detail the nature of angular momentum loss in stars as functions of mass and age. For example, we find that the time-derivative of the angular momentum distribution exhibits a clear inflection at the fully convective boundary, suggesting that the dominant angular momentum loss mechanism(s) may be qualitatively different for fully convective stars and not merely quantitatively weaker.

The typical scatter in the resulting gyrochronology-based age estimates is $\approx$0.2--0.3~dex, with the largest fractional uncertainties at the youngest ages where the intrinsic scatter in the rotation periods is a much larger fraction of the average. 

Variable stars with radiative envelopes also have some sensitivity to age (see Appendix~\ref{sec:radiative}). We find a clear separation in primary periods between main sequence and subgiant $\delta$~Scuti stars in this age limited sample. Similarly, we find a slight age gradient in slowly pulsating B-type variables, with the youngest stars stars having shorter periods. In $\gamma$~Dor type variables there does not appear to be a strong link between the dominant period and their age, however, their pulsation period does appear to be related to their rotation period, making it possible to extend the analysis of angular momentum evolution beyond the Kraft break. In the future, as all these pulsating variables tend to be highly periodic, it may be possible to further analyze higher pulsation frequencies to better understand their dependence on age.

These data represent an important step forward in understanding rotation period and angular momentum evolution of stars as a function of age. No longer limited to just a handful of canonical clusters, angular momenta can now be measured for a large number of populations with a near-continuous age distribution. Such an unprecedented data volume allows for the most comprehensive analysis of gyrochronology relations to date, making it possible to estimate ages of field stars with greater precision.



\appendix 

\section{Periodic variables among stars with radiative envelopes} \label{sec:radiative}

Sources bluer than the Kraft break in Figure \ref{fig:periodogram} tend to be concentrated in different parts of the color/period parameter space. In this section we provide a brief overview of the different variable types that inhabit this parameter space. They are included primarily for identification purposes, but their detailed analysis is beyond the scope of this work.

\subsection{$\delta$ Scuti}

$\delta$ Scuti variables have periods shorter than 8 hours days \citep[e.g.,][]{breger2000,handler2009,balona2015}. While typically the observed distribution of periods in literature is continuous for such variables, in this sample we observe strong bimodality - high frequency sources with the primary period shorter than 1.2 hours, and low frequency sources with periods longer than 2 hours. There is a difference of these two types of $\delta$ Scuti variables on the HR diagram. High frequency sources tend to be located on the main sequence, and they can include stars of all ages. Low frequency sources tend to be older subgiants. As the sample is age-limited in comparison to the field, it is likely that the subgiants older than $>1$ Gyr would fill in this gap. This is consistent with the expectations of the period-luminosity relation \citep[e.g.,][]{ziaali2019}.

As the cadence in TESS FFI images used in this analysis is $\sim$30 minutes, and the minimum period that we go down to in deconvolving periodicity is 40 minutes, there may be some confusion regarding the true dominant period of the main sequence $\delta$ Scuti variables, as the beat period is highly pronounced. Their overall placement of having a dominant period$<$1.2 days, however, is certain. But, given that $\delta$ Scuti stars themselves tend to be highly multi-periodic, more finely-sampled observations in the future are needed to perform a detailed analysis of such star

However, using more finely sampled data, it may be possible to have a more comprehensive periodicity measurements down to much higher frequencies, unaffected by the systematics. In doing so, it may be possible possible to independently establish their ages via astroseismology \citep{bedding2020,pamos-ortega2022}.

\subsection{$\gamma$ Dor}
A particularly notable concentration is found in a narrow range of colors corresponding to late A/early F spectral types, with typical periods of 0.3--1.5 days. They are somewhat redder than what is observed in $\delta$ Scuti-type stars. These sources span full range of ages in our sample, although only stars still on the main sequence can be detected as variable in this parameter space. These properties suggest that these are most likely to be $\gamma$ Dor type stars \citep{pollard2009}, a subset of pulsating variables that experience non-radial g-type pulsations.

The rotation period and the pulsation frequency are thought to be coupled in these types of stars \citep{dupret2007}. While we can't detect their rotation directly due to lack of convective atmospheres leading to a lack of spots, this seems likely, as these $\gamma$ Dor variables being found just blueward of the Kraft break have very similar periods to the highest mass stars with convective envelopes found redward of the gap. However, no age gradient is apparent in their period distribution.

\subsection{SPB}

There is a concentration of massive variable stars spanning a range of colors consistent with mid-to late B stars. The shortest period sources ($\sim$0.3 days) tend to be very young, with a typical age of $\sim$5 Myr. Sources with a period of $\sim$0.5 days tend to be younger than 30 Myr. Even at the longer periods, few stars have ages older than 200 Myr.

These colors and period ranges are most consistent with slowly pulsating B-type (SPB) variables \citep{de-cat2002a}. However, such an age gradient has not been observed among them in the past.

\software{TOPCAT \citep{topcat}, \texttt{eleanor} \citep{feinstein2019}, \texttt{GaussPy} \citep{lindner2015,lindner2019}}

\begin{acknowledgments}

This work has made use of data from the European Space Agency (ESA)
mission {\it Gaia} (\url{https://www.cosmos.esa.int/gaia}), processed by
the {\it Gaia} Data Processing and Analysis Consortium (DPAC,
\url{https://www.cosmos.esa.int/web/gaia/dpac/consortium}). Funding
for the DPAC has been provided by national institutions, in particular
the institutions participating in the {\it Gaia} Multilateral Agreement.
\end{acknowledgments}

\bibliographystyle{aasjournal.bst}

\end{document}